\pdfoutput=1
\documentclass[preprint]{sig-alternate-05-2015}
\usepackage{balance,xspace,fancyhdr}
\usepackage[inline]{enumitem}
\usepackage{graphicx}
\usepackage{hyperref}
\hypersetup{
	hidelinks  = true,
	colorlinks = true,
	linkcolor  = black,
	citecolor  = blue,
	urlcolor   = black,
}
\usepackage{cleveref}
\usepackage{siunitx}
\usepackage{multicol, multirow, bigstrut}
\usepackage{adjustbox}
\usepackage{color}
\usepackage{amsmath}
\usepackage{relsize}
\usepackage{flexisym}
\usepackage{makecell}
\usepackage{colortbl}
\usepackage{tikz}
\usetikzlibrary{arrows}
\usepackage[space, sort]{cite}
\usepackage{mathtools}
\usepackage{enumitem}

\newcommand\topStrut{\rule{0pt}{2em}}

\newcommand{\leftspecialcell}[2][c]{\begin{tabular}[#1]{@{}l@{}}#2\end{tabular}}
\newcommand{\centerspecialcell}[2][c]{\begin{tabular}[#1]{@{}c@{}}#2\end{tabular}}

\newcommand{\ruledef}[2]{$\dfrac{\begin{array}[c]{c}#1\end{array}}{\begin{array}[c]{c}#2\end{array}}$}
\newcommand{\rulename}[1]{\relsize{0}{\color{gteal}[\textsc{#1}]}}

\newcommand{\norm}[1]{\textnormal{#1}}

\newcommand{\isubtype}[2]{#1<:#2}
\newcommand{\ifamily}[2]{#1\ll: #2}
\newcommand{\phd}{{\,\_}}
\newcommand{\pts}{{pts}}
\newcommand\ripple{\textsc{Ripple}\xspace}
\newcommand\strinf{\textsc{StrInf}\xspace}
\newcommand\elf{\textsc{Elf}\xspace}
\newcommand\solar{\textsc{Solar}\xspace}
\newcommand\droidRA{\textsc{DroidRA}\xspace}
\newcommand\spark{\textsc{Spark}\xspace}
\newcommand\soot{\textsc{Soot}\xspace}
\newcommand\FlowDroid{\textsc{FlowDroid}\xspace}
\newcommand\checker{\textsc{Checker}\xspace}

\definecolor{gteal}{RGB}{255, 0, 0}
\definecolor{gblue}{RGB}{0, 0, 255}
\definecolor{gray}{RGB}{96, 125, 139}
\definecolor{arrowred}{rgb}{0.961, 0, 0.341}
\definecolor{arrowblue}{rgb}{0.161, 0.384, 1}
\definecolor{redfill}{rgb}{0.980, 0.501, 0.671}

\newcommand\fauxsc[1]{\fauxschelper#1 \relax\relax}
\def\fauxschelper#1 #2\relax{%
  \fauxschelphelp#1\relax\relax%
  \if\relax#2\relax\else\ \fauxschelper#2\relax\fi%
}
\def\Hscale{.80}\def\Vscale{.72}\def\Cscale{1.0}
\def\fauxschelphelp#1#2\relax{%
  \ifnum`#1>``\ifnum`#1<`\{\scalebox{\Hscale}[\Vscale]{\uppercase{#1}}\else%
    \scalebox{\Cscale}[1]{#1}\fi\else\scalebox{\Cscale}[1]{#1}\fi%
  \ifx\relax#2\relax\else\fauxschelphelp#2\relax\fi}

\makeatletter
\@ifundefined{thepage}{\def\thepage{\arabic{page}}}{}%
\makeatother

\newcommand*{\refname}{Bibliography}

\begin{document}

\title{
	Ripple: Reflection Analysis for Android Apps in \\
 Incomplete Information Environments\\
}

\numberofauthors{1}
\author{
\alignauthor
	Yifei Zhang, Tian Tan, Yue Li and Jingling Xue \\
	\affaddr{Programming Languages and Compilers Group} \\
	\affaddr{School of Computer Science and Engineering, UNSW Australia} \\
	\email{\{yzhang, tiantan, yueli, jingling\}@cse.unsw.edu.au}
}

\maketitle
\begin{abstract}
Despite its widespread use in Android apps,
reflection poses graving problems for static 
security analysis. Currently, string inference 
is applied to handle reflection,
resulting in significantly missed security 
vulnerabilities. In this paper, we bring forward
the ubiquity of
incomplete information environments (IIEs)
for Android apps, 
where some critical data-flows 
are missing during static analysis, and 
the need for resolving reflective calls under IIEs.
We present \ripple, the first IIE-aware static reflection analysis for Android apps that resolves reflective calls more soundly than 
string inference. Validation with 17 popular Android apps from Google Play demonstrates the effectiveness of \ripple in 
discovering reflective targets with a
low false positive rate. As a result, \ripple enables FlowDroid, a taint analysis for Android apps, to find hundreds of sensitive data leakages that would otherwise be missed. 
As a fundamental analysis, \ripple will
be valuable for 
many security analysis clients, since
more program behaviors can now be analyzed under IIEs.
\end{abstract}

%
%

%
%

%
%



\section{Introduction}

Android's increase in popularity 
and its openness have triggered a great rise in malware-spreading apps.
Static analysis is a
fundamental tool for detecting security threats
in Android apps early at software development time
at significantly reduced cost.
This approach is immune to emulation detection and
can provide a comprehensive picture of an app's 
possible behaviors. Therefore, static analysis finds 
diverse applications, including
data leakage detection \cite{flowdroid, apposcopy, droidsafe, hybridroid}, repackaging attack 
detection~\cite{Zhou12,chen2014achieving}, security policy verification \cite{checker,sparta,
rasthofer2014droidforce}, security vetting \cite{chex, amandroid,Grace12}, privacy violations~\cite{
slavin2016toward}, and malware detection \cite{drebin,Zheng13,profiling}. 

However, reflection, under which a class, method or field
can be manipulated by its string name,
poses graving problems for
static analysis. 
According to a recent
study~\cite{Rastogi13}, malware authors can use
reflection to hide malicious behaviors from detection by
all the 10 commercial static analyzers tested. 
Similarly, academic static analyzers either ignore
reflection \cite{triggerscope,ic3,epicc,
describeme} or handle 
it only partially~\cite{flowdroid}, 
resulting in also significantly missed
program behaviors. 

Reflection analysis aims to discover
statically the reflective targets referred to 
at reflective calls. For Android apps,
regular string inference is currently performed
to discover the string constants used as
class/method/field names at reflective calls 
\cite{droidra,checker,sparta}.
For example, if 
\texttt{cName} in \texttt{clz = Class.forName(cName)} is 
statically discovered to be ``\texttt{A}'', then we know
that \texttt{v} points to an object 
of type \texttt{A} reflectively created at 
\texttt{v = clz.newInstance()}. If
\texttt{cName} is statically unknown, then  
\texttt{v = clz.newInstance()} is ignored.

Regular string inference is inadequate for
framework-based and event-driven Android apps.
In practice, reflection analysis must be performed
together with many other analyses, including pointer
analysis 
\cite{elf, solar, yannisreflection, livshits,droidsafe},
inter-component communication (ICC)
analysis~\cite{epicc, ic3, primo},
callback analysis \cite{flowdroid, droidsafe, gator, 
amandroid}, and
library summary generation~\cite{stubdroid, aikenmodeling, droidsafe}. Soundness, which demands 
over-approximation, is often sacrificed in order 
to achieve efficiency and precision tradeoffs.
As a result, class/method/field names used at
reflective calls may be non-constant (either non-null but
statically unknown or simply null). Similarly, 
the receiver objects at reflectively method call sites 
(i.e., the objects pointed to by \texttt{v} in
\texttt{Method.invoke(v,\ldots)})  
may be non-null but with statically unknown types or simply null.
Such information
can be missing due to, for example, unsound library summaries, 
unmodeled Android services, code obfuscation,
and unsound handling of hard-to-analyzed Android 
features such as ICC, callbacks and built-in 
containers. In this case, regular string inference, which keeps track
of only constant strings, is ineffective, resulting in missed program
behaviors.

In this paper,
we bring forward the ubiquity of
incomplete information environments (IIEs) for Android
apps, where some critical data-flows 
are inevitably missing during static analysis. As discussed above,
these include not only the case 
when class/method/field names are non-null but statically unknown,
which is studied previously for 
Java programs~\cite{elf, solar, yannisreflection, livshits}, 
but also the case when these
string names are null, which is investigated for the first time
for Android apps in this paper. We therefore 
emphasize the need for resolving reflective calls 
in Android apps under IIEs.
To this end,
we introduce \ripple, the first IIE-aware static reflection analysis for Android apps that
can resolve reflective calls more soundly than string inference 
at a low false positive rate.
We also demonstrate its effectiveness in improving the
precision of an important security analysis.

In summary, this work makes the following contributions:
\begin{itemize}[topsep=1pt, itemsep=-2pt]
\item We present (for the first time) an empirical study for IIEs in
real-world Android apps, and examine some
common sources of incomplete information, 
discuss their impact on reflection analysis, and
motivate the need for developing an IIE-aware reflection
analysis.

\item We introduce \ripple, the first IIE-aware
reflection analysis for Android apps, which performs
type inference automatically (without requiring 
user annotations) and thus subsumes regular string inference.

\item
We have implemented \ripple in \soot \cite{soot}, a 
static analysis and optimization framework for Java
and Android programs, with \ripple working together
with its \spark, a flow- and context-insensitive 
pointer analysis. 
We have evaluated the soundness, precision, scalability
and effectiveness of \ripple by using 17 popular 
real-world Android apps from Google Play, in which the
data-flows needed for resolving some reflective calls are null.
\ripple discovers 72 more (true)
reflective targets than string inference in seconds
at a low false positive rate of 21.9\%.
This translates into 310 more sensitive data leakages 
detected by 
\FlowDroid \cite{flowdroid}. 
\end{itemize}

The rest of this paper is organized as follows.
Section~\ref{sec:back} reviews reflection usage
and discusses reflection-related security 
vulnerabilities.
Section~\ref{sec:iie} examines the ubiquity of IIEs.  
Section~\ref{sec:meth} describes our 
\ripple approach.
Section~\ref{sec:form} gives a formalization of 
\ripple.
Section~\ref{sec:eval} evaluates \ripple with
real-world Android apps.
Section~\ref{sec:rel} discusses the related work.
Finally, 
Section~\ref{sec:conc} concludes.

\section{Background}
\label{sec:back}

We review reflection usage in Android apps and
discuss security issues caused if reflection
is not handled well.

\subsection{Reflection Usage}

Android apps are coded in Java. The Java reflection 
API provides metaobjects for inspecting
classes, methods, and fields at runtime.

\begin{figure}[t]
\vspace*{1ex}
\centering
\includegraphics[width=\linewidth]{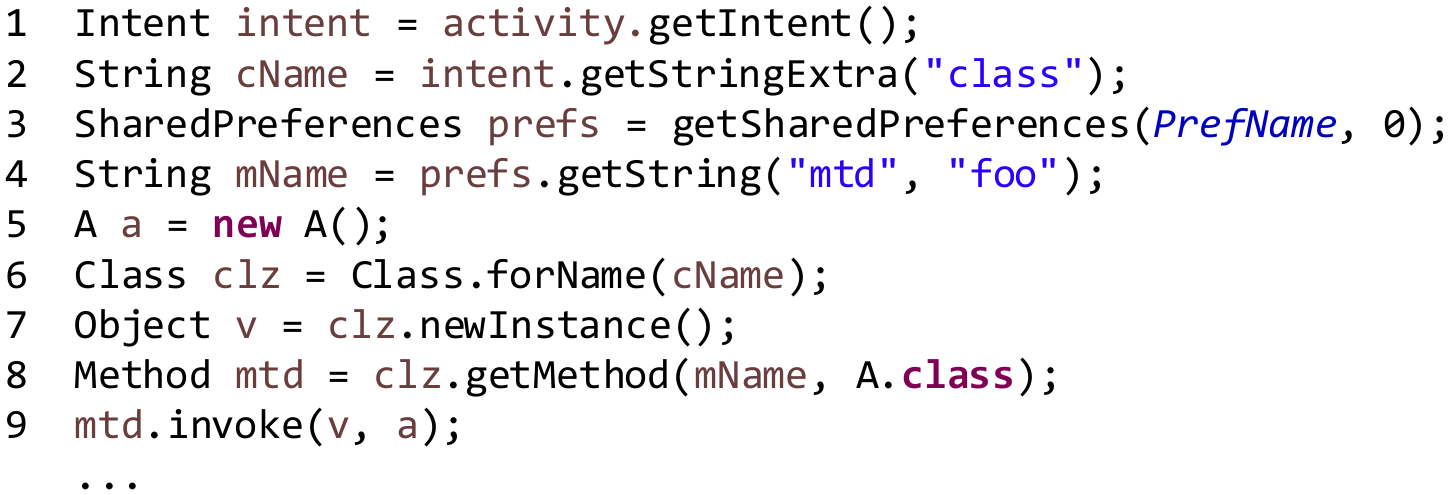}
\vspace*{-4ex}
\caption{An example of reflection usage in Android.}
\label{background:example}
\end{figure}

In \Cref{background:example}, \texttt{clz} and 
\texttt{mtd} are metaobjects of classes 
\texttt{Class} and \texttt{Method}, respectively. 
Their names are obtained from \texttt{Intent} and 
\texttt{SharedPreferences}, two Android's
built-in containers. In lines 1 -- 2,
the class name \texttt{cName} for \texttt{clz} is 
retrieved from an intent. Subsequently,
in line 6, \texttt{clz} is created in a call to 
\texttt{Class.forName()}.
In line 7, \texttt{v} points to
an object reflectively created by 
\texttt{clz.newInstance()}. In lines 3 -- 4, 
the method name
for \texttt{mtd} is retrieved from the map \texttt{prefs}  as the value associated
with the key ``\texttt{mtd}'' or ``\texttt{foo}'' if the key 
``\texttt{mtd}'' 
does not exist yet. In line 8, \texttt{mtd} is created
in a call to 
\texttt{clz.getMethod()}, which returns  
a public method declared in or inherited by \texttt{clz} with a single
parameter of type \texttt{A}. Finally,
in line 8, this method is called reflectively
by \texttt{mtd.invoke(v, a)} on the receiver object 
pointed by \texttt{v} with the actual argument being
\texttt{a}. If \texttt{mtd} is
static, then \texttt{v} is null.

Reflection introduces implicitly the caller-callee
edges into the call graph of the program. If one
reflective call, e.g.,
\texttt{Class.forName()},
\texttt{clz.getMethod()},
\texttt{clz.newInstance()} or
\texttt{mtd.invoke()}, is ignored, the caller-callee
edges in line 9 will not be discovered. As a result,
possible security vulnerabilities in the invisible
part of the program will go undetected.

Therefore, the objective of reflection analysis is to
discover the targets at reflective calls (e.g., 
objects created, methods called and
fields accessed), by working with pointer analysis.  

\subsection{Reflection-Related Security Issues}

In Android apps,
reflection serves a number of purposes, including 
(1) plug-in and external library support,
(2) hidden API method invocation,
(3) access to private API methods and fields, 
and (4) backward compatibility. Indeed,
reflection is widely used in both benign and malicious 
Android apps. 
For a sample
of 202 top-chart free apps from Google Play that
we analyzed on 
15 April 2016, we found that 92.6\% of
these apps use reflection. Elsewhere, in a malware sample consisting of
\num{6141} Android apps from the VriusShare project \cite{virusshare}, 48.13\% of them also use reflection.

Reflection is responsible for a number of
security exploits in Android apps. In general, 
a malicious app retrieves class and method 
names as strings externally and invokes methods in
payload classes
via reflection to perform malicious activities, which
are thus disguised from some signature-based anti-virus
software. 
For example, Obad \cite{obad} and FakeInstaller \cite{fakeinstaller}, represent the two
most sophisticated malware families~\cite{harvester},
as they combine reflection and code obfuscation to 
hide their malicious behaviors. In both cases, 
the methods that are used to collect and steal 
sensitive data 
are invoked reflectively with encrypted class and 
method names.
To elude detection by dynamic analyzers such as 
Google Bouncer \cite{bouncer}, 
malicious behaviors are also suppressed by using
logic bombs~\cite{triggerscope} and emulation
detection mechanisms. 

As a result, effective static analysis for reflection
with a good precision is needed by the analysis community
for Android apps. With such a tool,
the malicious behaviors in malware and 
the security vulnerabilities caused by misused 
reflection in goodware can both be detected.

\section{IIEs in Android apps}
\label{sec:iie}
\label{sec:motivating}

There are two types of missing information under IIEs. In one
case, the data flows needed for resolving reflective calls 
exist but are statically unknown. Consider the code given
in \Cref{background:example}. During the static analysis,
\texttt{cName} and 
\texttt{mName} may be non-null but are statically unknown.
Similarly, \texttt{v} may point to a non-null object but with a
statically unknown type. In this case, we can resolve
the reflective calls in lines 6 -- 9 by performing type
inference to infer what \texttt{clz},
\texttt{mtd}, and the objects pointed to by  \texttt{v} are, as
done previously for Java
programs~\cite{elf, solar, yannisreflection}.

In the other case, the data flows needed for resolving reflective 
calls are completely missing, indicated by the presence of null.
To understand this case, 
which has never been studied before,
for Android apps, we have performed
an empirical study on 45 Android apps, with 
20 popular Android apps from Google Play and 
25 malware samples from the
VirusShare Project~\cite{virusshare}. 
We discuss the four most common sources of incomplete
information in IIEs: (1) undetermined intents, 
(2) behavior-unknown libraries, (3)
unresolved built-in containers,
and (4) unmodeled services. 
We delve into their bytecode to explain why  regular string
inference is inadequate, since it fails   to enable
\FlowDroid~\cite{flowdroid} to discover many data leaks from
\emph{sources} (API calls that inject sensitive information) 
to \emph{sinks} (API calls that leak information).
We also provide insights
on why our IIE-aware \ripple
can handle IIEs more effectively than string inference. 

\subsection{Undetermined Intents}

ICC via intents is one of the most fundamental 
features in Android as it enables some 
components to process the data originating from
other components. Thus, the components in
an Android app function as building blocks for
the entire system, enhancing intra- and 
inter-application code reuse.

\begin{figure}[h]
\includegraphics[width=\linewidth]{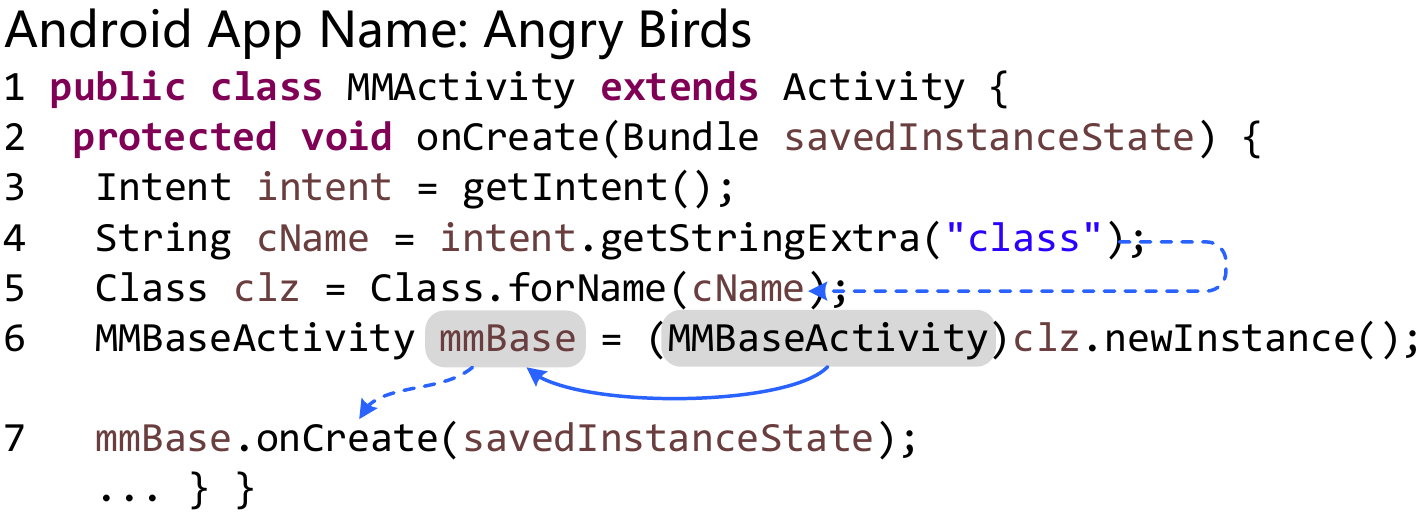}
\caption{Undetermined intents.
Here, \protect \tikz[baseline=-0.5ex] \protect \draw [densely dashed, line width = 0.8pt, draw = arrowblue, arrows = {-latex}, fill = arrowblue] (0, 0) -- (0.7, 0);
denotes a missed data-flow and
\protect \tikz[baseline=-0.5ex] \protect \draw [line width = 0.8pt, draw = arrowblue, arrows = {-latex}, fill = arrowblue] (0, 0) -- (0.7, 0);
the post-dominating-cast-based type inference used in \ripple. These notations are also used in Figures~\ref{motivating:library} -- \ref{motivating:framework}.
}
\label{motivating:intent}
\end{figure}

In practice,
some inter-component control- and data-flows
cannot be captured by ICC 
analysis~\cite{epicc, ic3, primo}. If a data flow
from an intent into 
a reflective call is missing, the reflection call
cannot be fully resolved. The code snippet in \Cref{motivating:intent} taken from the game \textit{Angry Birds} illustrates this problem.

In this app, the class name \texttt{cName}
is obtained from an intent (line 4) and then
used in a call to 
\texttt{Class.forName()} (line 5) to create
a class metaobject \texttt{clz}. Then, an
object of this class is created reflectively (line 6)
and assigned to \texttt{mmBase} after a downcast to
\texttt{MMBaseActivity} is performed. Finally, 
\texttt{onCreate()} is invoked on this object (line 7).

To discover what \texttt{cName} is, we applied IC3,
a state-of-the-art ICC analysis~\cite{ic3}, but
to no avail. Thus, the data-flow for \texttt{cName}, 
denoted by 
\tikz[baseline=-0.5ex] \protect \draw [densely dashed, line width = 0.8pt, draw = arrowblue, arrows = {-latex}, fill = arrowblue] (0, 0) -- (0.5, 0);,
is missing, rendering \texttt{clz} to be 
a null pointer.  In this case, string inference is ineffective.
As a result,
the reflectively allocated object in line 6 and the
subsequent call on this object in line 7 are ignored.

\ripple is aware of the incomplete information
caused by this undetermined intent, which manifests
itself in the form of
\texttt{cName = null}. By taking advantage
of the post-dominant cast \texttt{MMBaseActivity}
for \texttt{clz.newInstance()} in line 6, \ripple
infers that \texttt{mmBase} may point to five
objects with their types ranging over
\texttt{MMBaseActivity} and its four subtypes, which
are all confirmed to be possible by manual code
inspection. As a result,
\ripple discovers 
\num{3928} caller-callee edges 
in lines 5 -- 7 (directly or indirectly), 
thereby enabling
\FlowDroid \cite{flowdroid} to detect
49 new sensitive data leaks that will be all missed
by string inference in this part of the app that
has been made analyzable by \ripple.

\subsection{Behavior-Unknown Libraries}
\label{sec:unknown-lib}

To accelerate the analysis of an application, the
side effects of a library on the application are
often summarized. Library 
summaries are either written manually~\cite{droidsafe}
or generated
automatically~\cite{stubdroid,aikenmodeling}.
However, both approaches are error-prone and often
fail to model all the side-effects of a library
for all possible analyses.
DroidSafe \cite{droidsafe} provides the
Android Device Implementation (ADI) to model the 
Android API and runtime manually, with about
1.3 MLOC for Android 4.4.3. However, as the
Android framework evolves with both new features 
and undocumented code added, how to keep this ADI 
in sync can be a daunting task.

\begin{figure}[h]
\includegraphics[width=\linewidth]{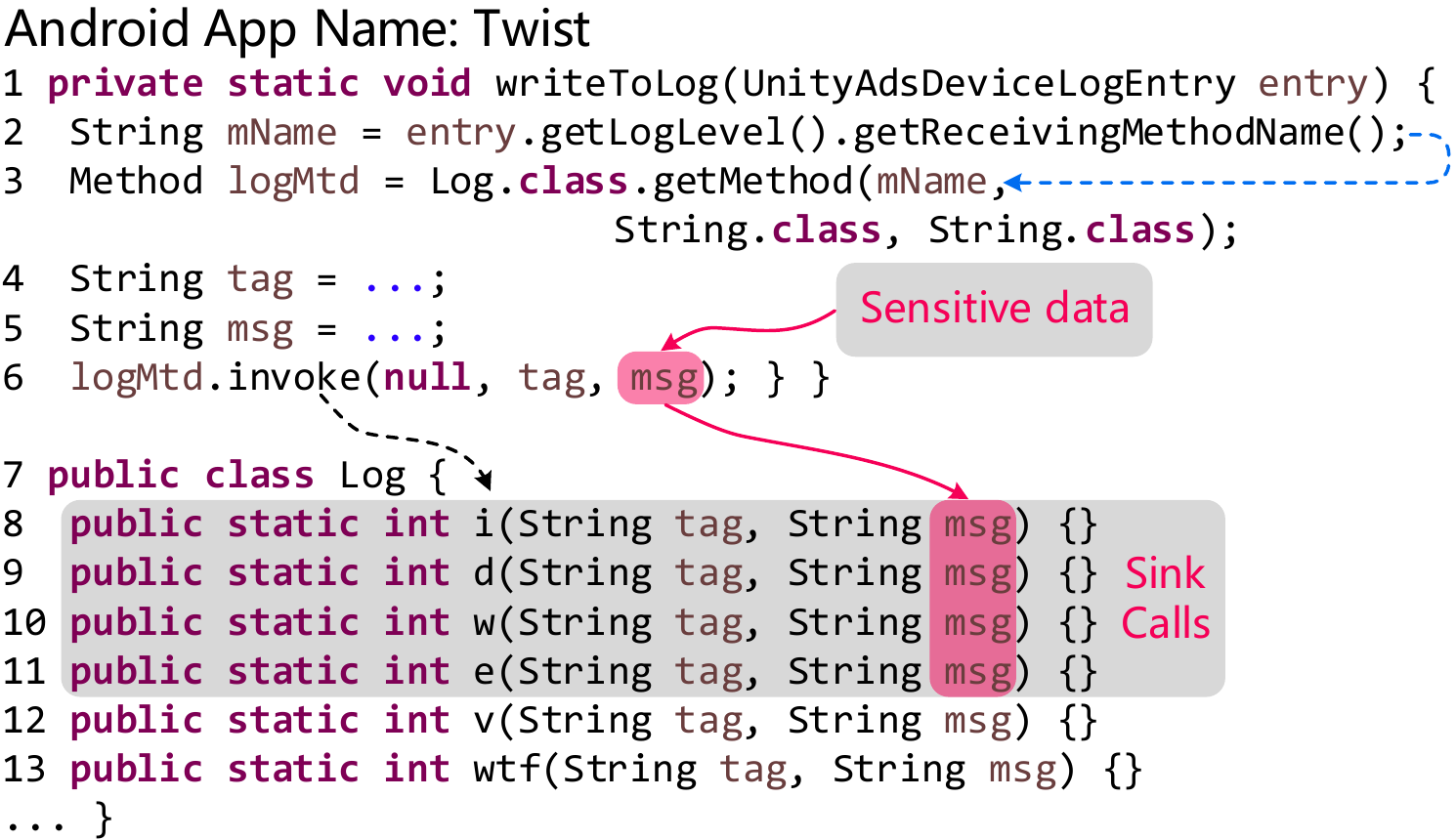}
\caption{Behavior-unknown libraries. Here,
\protect \tikz[baseline=-0.5ex] \protect \draw [densely dashed, line width = 0.8pt, draw = black, arrows = {-latex}] (0, 0) -- (0.7, 0);
marks the target methods invoked
at a reflective call,
\protect \tikz[baseline=-0.5ex] \protect \draw [line width = 0.8pt, draw = arrowred, arrows = {-latex}, fill = arrowred] (0, 0) -- (0.7, 0);
denotes sensitive data-flow, and
\protect \tikz \protect \draw [rounded corners, redfill, fill = redfill] (-.1, -.2) rectangle(0.4, 0.);
	denotes tainted data. These notations,
	together with those in Figure~\ref{motivating:intent},
	are also used in Figures~\ref{motivating:bundle} and \ref{motivating:framework}.}
	\label{motivating:library}
\end{figure}

Therefore, unsound library summaries
are an important source of incomplete
information in IIEs.
The code snippet in \Cref{motivating:library} taken 
from the app \textit{Twist} illustrates this issue.

This code snippet is used to log messages
at different verbosity levels.  In line 2,
a method name \texttt{mName} is retrieved. In line 3,
its method metaobject \texttt{logMtd}
is created. In
line 6, this method, which is static, is invoked
reflectively.

If we apply \FlowDroid \cite{flowdroid} to
detect data leaks in this app, by relying on string inference to
perform reflection analysis,
then the reflective call  \texttt{logMtd.invoke()} in
line 6 will
be ignored. In \FlowDroid, the behaviors of maps
are not summarized. However, \texttt{entry}
was retrieved from a \texttt{HashMap}
and then passed to \texttt{writeToLog}. Thus,
\texttt{mName = null}, rendering string inference
to be ineffective.

\ripple is aware of 
unsound library summaries
and thus attempts to infer the target methods at 
\texttt{logMtd.invoke()}. Based on the facts that (1)
these methods are static (since the receiver object
is null), declared in or inherited 
by class \texttt{android.util.Log},
(2) each target method has two formal parameters, and 
(3) each parameter has a type that is either
\texttt{String} or its supertype or its subtype, \ripple
concludes that the six target methods,
\texttt{i()},
\texttt{d()},
\texttt{w()},
\texttt{e()},
\texttt{v()} and
\texttt{wtf()},
as shown
in class \texttt{android.util.Log} may be potentially
invoked. According to \FlowDroid, these six
methods are all sinks for sensitive data 
contained in \texttt{msg}.
Thus, resolving 
\texttt{logMtd.invoke()} causes
12 data leaks from two different sensitive data 
sources to be 
reported (as 2 sources $\times$ 6 sinks = 12 leaks). 
By manual code inspection,
we found that the first four methods,
\texttt{i()},
\texttt{d()},
\texttt{w()} and
\texttt{e()}, shaded
in class \texttt{android.util.Log} are
true targets, implying that 8 data leaks will not be
reported if string inference is used.

\subsection{Unresolved Built-in Containers}

\begin{figure}[h]
	\includegraphics[width=\linewidth]{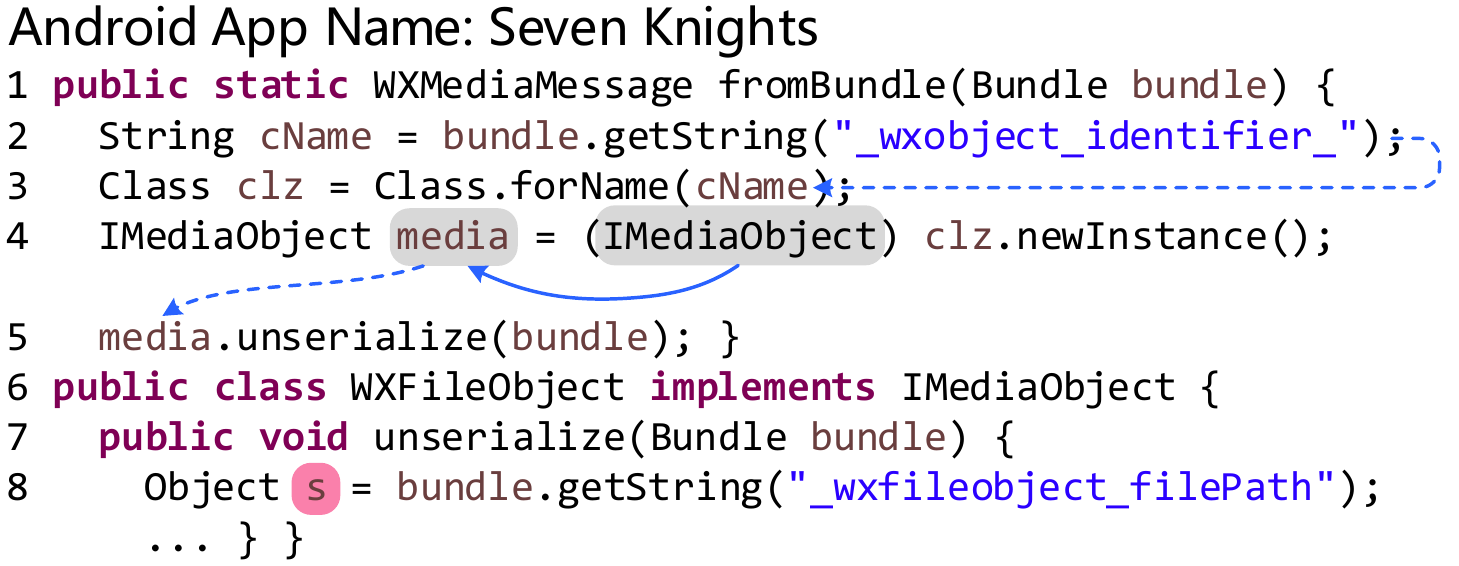}
	\caption{Unresolved Bundles.}
	\label{motivating:bundle}
\end{figure}

Android apps can receive a variety of user inputs from,
e.g., intents, databases, internet, GUI actions, and 
system events. These data are stored in different types of containers, such as 
\texttt{Bundle}, \texttt{SharedPreferences}, \texttt{ContentValues} and \texttt{JSONObject}, for different 
purposes. 
Unhandled user inputs represent an important source of 
incomplete information in IIEs.
The code snippet in \Cref{motivating:bundle} taken from a game named \textit{Seven Knights} illustrates this problem.

In this code snippet, different types of objects are 
created (line 4) to handle different types of
media data according to their unique identifiers 
stored in a \texttt{Bundle} (line 2), which 
is constructed according to the types of media 
introduced by third-party apps. 
\texttt{IMediaObject} is an interface implemented 
by eight types (i.e., classes) of media, with only
\texttt{WXFileObject} 
shown partially. Therefore, \texttt{cName} represents
the name of one of these eight classes.
In line 4, \texttt{media} points
to a reflectively created object of the class 
identified by \texttt{cName}. In line 5, a call is
made to 
\texttt{unserialize()} on the receiver object
pointed to by \texttt{media} with \texttt{bundle}
as its argument.

If we apply again \FlowDroid to detect data leaks
in this app, by relying on  string inference
to resolve the reflective calls
in the app, then \texttt{cName} will be null, since 
the behaviors of bundles are not modeled in
\FlowDroid. As a result, the reflectively created
object in line 4 and the subsequent call on this
object in line 5 will be ignored.

By being IIE-aware, \ripple
will infer the inputs retrieved from \texttt{bundle}
to resolve the call to \texttt{clz.newInstance()} in
line~4. By taking advantage of the post-dominant cast 
\texttt{IMediaObject} for this reflective 
call, \ripple deduces that
\texttt{media} points to potentially
eight objects with their types ranging over 
all the eight classes implementing
\texttt{IMediaObject}, confirmed by manual code 
inspection. As a result, a total of
37 caller-callee edges, together with 16 sensitive
data sources, which would otherwise be missed
by string inference, are discovered in lines 3 -- 5
directly or indirectly. Currently, these 16 sensitive
data sources do not flow to any sinks but may do so 
in a future app release. The resulting leaks will
be then detected by \FlowDroid, assisted by \ripple.

\subsection{Unmodeled Services}

\begin{figure}[h]
	\vspace*{1ex}
	\includegraphics[width=\linewidth]{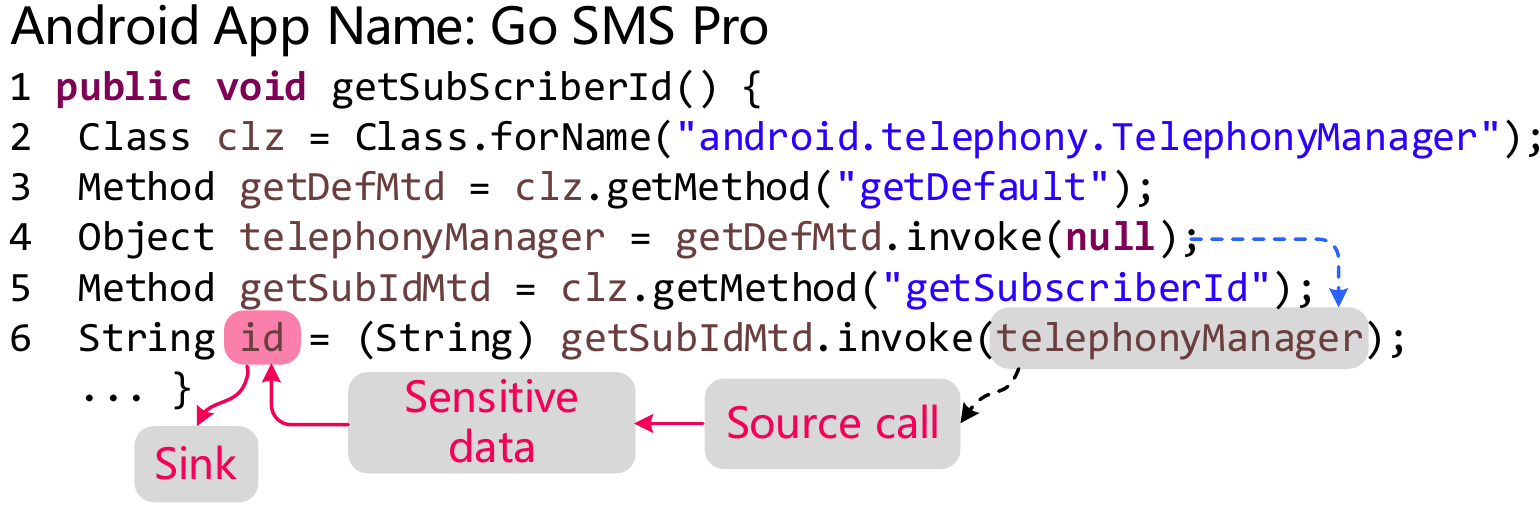}
	\caption{Unmodeled services.}
	\label{motivating:framework}
\end{figure}

The Android framework provides an abstraction
of abundant services for a mobile device, such as 
obtaining the device status, making phone calls, and
sending text messages, which are all related to 
critical program behaviors. These services are usually 
initialized during system startup and subsequently
used by calling the factory methods in the
Android framework with often reflective calls involved. 
Unsound modeling for
Android's system-wide services can be
an important source of incomplete information in IIEs. 
The code snippet in \Cref{motivating:framework} taken from a text message management app named \textit{GO SMS Pro} illustrates this issue.

In line 2, \texttt{clz} represents a class metaobject
for \texttt{android.\linebreak telephony.TelephonyManager}.
In line 3, \texttt{getDefMtd} represents a method
metaobject for a 
static method named \texttt{getDefault} in \texttt{clz}.
In line 4, this method is invoked reflectively,
with its
returned object, an instance of \texttt{clz},
assigned to \texttt{telephonyManager}.
In lines 5 -- 6, a method metaobject, 
\texttt{getSubIdMtd}, for an instance
method named \texttt{getSubscribeId} in \texttt{clz} 
is created and then invoked reflectively on the
receiver object pointed to by \texttt{telephonyManager}.

In this code snippet, all the class and method names
are string constants. Thus, regular string inference
can resolve precisely the reflective targets at all the
reflective calls shown.
However, this still does not enable 
the target methods invoked 
in line~6 to be analyzed, because the
\texttt{getDefault} method invoked in line 4 is part
of the hidden API and thus not available for analysis. 
Thus, \texttt{telephonyManager} is null, causing the
reflective call in line 6 to be skipped.

\ripple is aware of the existence of unmodeled services.
By examining the class type in the \texttt{getSubIdMtd}
metaobject, \ripple concludes that \texttt{telephonyManager} points to an object of 
type \texttt{android.telephony.TelephonyManager}. As
a result, the reflective call in line 6 can be 
resolved, resulting in  the target method
\texttt{getSubscriberId} to be discovered. For this
app, \FlowDroid is unscalable. Otherwise, the potential
data link as shown will be detected automatically.

Finally, if the \texttt{getDefault} method is native
with its method body unmodeled but available for
analysis, then \ripple will be able to infer in line 4
that 
\texttt{telephonyManager} may point to an object of 
type \texttt{android.telephony.TelephonyManager}. 
As a result, the reflective call in line 6 can also
be resolved.

\section{Methodology}
\label{sec:meth}

\begin{figure*}[ht]
	\centering
	\includegraphics[width=0.95\linewidth]{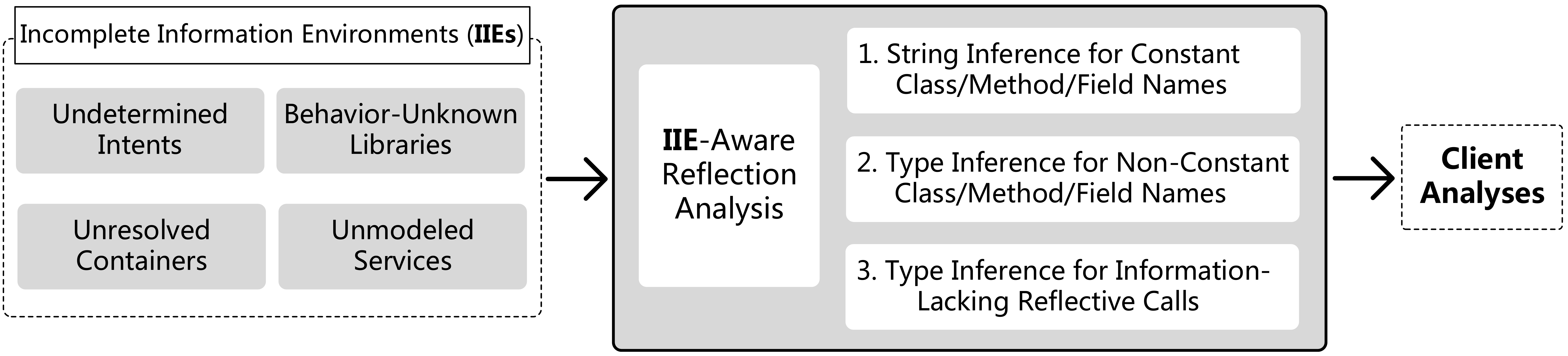}
	\caption{An overview of \protect\fauxsc{Ripple}.}
	\label{method:overview}
\end{figure*}

\Cref{method:overview} depicts an overview of \ripple,
an IIE-aware reflection analysis introduced in this paper
for Android apps. 
Currently, 
we consider four important
contributing factors to IIEs when the data-flows
needed for resolving reflective calls are null:
undetermined intents, behavior-unknown libraries, 
unresolved containers and unmodeled services,
as discussed in Section~\ref{sec:iie}. 
To handle IIEs effectively, \ripple resolves reflective
calls in the presence of incomplete information about 
these calls, so that their induced caller-callee edges 
can be discovered. Once the call graph of an app is
available, many client analyses,
such as data leakage detection, security policy verification, malware detection, and security vetting, can be
performed to detect various security issues, especially
those hidden in some reflective calls, in the app.

We have developed \ripple by leveraging 
recent advances on reflection analysis for 
Java~\cite{elf, solar}. Conceptually,
\ripple performs reflection analysis by distinguishing
three cases:
\begin{itemize}[leftmargin=*]
\item {\bf Case 1. String Inference for Constant Strings.} 
If class/method/field names used at reflective
calls
are string constants, 
regular string inference is conducted.

\item {\bf Case 2. Type Inference for Unknown Strings.} 
If class/method/field names
are non-constant but non-null strings, which may be
read from configuration files or command lines, 
then type inference that was previously introduced for 
Java programs~\cite{elf, solar, yannisreflection, livshits} can
be leveraged.

\item {\bf Case 3. Type Inference for Information-Missing
	Reflective Calls.} Again, type inference is
performed to infer the missing information at 
reflective calls in the following
three categories, as reviewed in Section~\ref{sec:iie}: 
\begin{itemize}[leftmargin=*]
\item {\bf Null-Name.} Class, method or field names
are null, as illustrated in 
Figure~\ref{motivating:intent}
(with \texttt{cName = null} for undetermined intents), 
Figure~\ref{motivating:library} (with
\texttt{mName = null} for behavior-unknown
libraries), and
Figure~\ref{motivating:bundle}
(with \texttt{cName = null} for unmodeled bundles).
We will replace a null object by an unknown
string and then go back to Case~2.

\item {\bf Missing-RecvObj.} 
Given a call to \texttt{mtd.invoke(y,\ldots)}, 
a target method pointed to by \texttt{mtd}
does not have a corresponding receiver object pointed to
by \texttt{y},
as illustrated in Figure~\ref{motivating:framework}
for the \texttt{getSubscriberId} method invoked in line
6, where \texttt{telephonyManager = null}. If we know 
the class type of the target method pointed to by 
\texttt{mtd}, its 
corresponding receiver object can be inferred.

\item {\bf Missing-RetObj.}
Given \texttt{x = mtd.invoke(\ldots)}, 
a target method pointed to by \texttt{mtd} is
available for analysis but its method body is 
unmodeled. This can happen to 
\texttt{telephonyManager = getDefMt.invoke(null)}
in Figure~\ref{motivating:framework},
as discussed in Section~\ref{sec:iie}, when
the \texttt{getDefault} method is hypothetically
assumed to be an unmodeled native method.
In general, if we know the return
type of the target method pointed to by \texttt{mtd}, 
then objects of this type or its subtypes
are created and assigned to \texttt{x}.

\end{itemize}
\end{itemize}

To the best of our knowledge, \ripple is the first automated
reflection analysis (without relying on
user annotations) for handling Cases 2 and 3 and
also the first for handling Cases 1 -- 3 in a unified framework
for Android apps.

In \ripple, reflection analysis is performed together
with pointer analysis mutually recursively,
as effectively one single analysis.
On one hand, reflection analysis makes use of the
points-to information to resolve reflective calls. On
the other hand, pointer analysis needs the results
of reflection analysis to determine which methods
get called and which fields get accessed. Once the
entire analysis for an app is over, its call graph is
readily available (at the same time).

\section{Formalism}
\label{sec:form}

We formalize \ripple as a form of reflection analysis,
performed together with a flow- and context-insensitive 
pointer analysis. We restrict
ourself to a small core of the Java reflection API,
including \texttt{Class.forName()}, 
\texttt{Class.newInstance()},
\texttt{Class.getMethod()}, and 
\texttt{Method.invoke()}.
For simplicity, we consider only instance methods as
static methods are handled similarly. We focus only
on reflective method calls as our formalism extends 
easily to reflective field accesses.

\subsection{Notations}

\Cref{formal:notation} gives the domains 
used in our formalism. The abstract heap 
objects are labeled by their allocation sites. 
$\mathbb{C}$ represents the set of class metaobjects
and $\mathbb{M}$ the set of method metaobjects.
The class type of a class or method
metaobject is identified by its superscript and
the signature of a method metaobject, which 
consists of the method name and descriptor (i.e.,
return type and parameter types), is identified by its
subscript. In particular,
$u$ indicates an unknown class type 
or an unknown method signature (with some parts of the
signature being statically unknown).

\begin{figure}[h]
	\hspace*{-1ex}
	\small
	\centering
	\begin{tabular}{ll}
		class type             & $ t, {u} \in \mathbb{T} $ \\
		variable               & $ v \in \mathbb{V} $ \\
		abstract heap object   & $ o^{t}_{1}, o^{t}_{2}, o^{\_}_{\_}, ... \in \mathbb{O} $\\
		class metaobject       & $ {c}^t, {c}^{u}, {c}^{\_}, ... \in \mathbb{C} $ \\
		method metaobject      & $ {m}^t_s, {m}^{u}_s, {m}^{t}_{u}, {m}^{u}_{u}, {m}^{\_}_{\_}, ... \in \mathbb{M} = \mathbb{T} \times \mathbb{S} $ \\
		method name            & $ n \in \mathbb{N} $ \\
		method parameter type  & $ p \in \mathbb{P} = \bigcup^{\infty}_{i=0} \mathbb{T}^i $ \\
		method signature       & $ s, {u} \in \mathbb{S} = \mathbb{T} \times \mathbb{N} \times \mathbb{P} $ 
	\end{tabular}
	\caption{Domains.}
	\label{formal:notation}
	\vspace*{-2ex}
\end{figure}

\subsection{Pointer Analysis}

\Cref{formal:pta} gives a standard formulation of
a flow- and context-insensitive Andersen's
pointer analysis~\cite{andersen1994program}.
$pts(v)$ represents the points-to set of a pointer $v$. 
An array object is analyzed with its elements collapsed 
to a single field, say, $arr$. 

\begin{figure}[ht]
	\centering
	\vspace*{1ex}
	 \relsize{-1}
\begin{tabular}{cc|cc}
\arrayrulecolor{gray}\hline
$i:$ \color{gblue} \text{x} = {\tt new} t() & \rulename{P-New} & 
\color{gblue}\text{x} = \text{y} & \rulename{P-Copy} \bigstrut \\ 

\multicolumn{2}{c|}{\ruledef{\phantom{abc}}{\{o^t_i\} \subseteq pts($x$)}} &
\multicolumn{2}{c}{\ruledef{\phantom{abc}}{pts($y$) \subseteq pts($x$)}}\bigstrut[b] \\[1em]

		\arrayrulecolor{gray}\hline
		\color{gblue}\text{x} = \text{y.f} & \rulename{P-Load} &
		\color{gblue} \text{x.f} = \text{y} & \rulename{P-Store} \bigstrut \\ 

		\multicolumn{2}{c|}{\ruledef{o^t_i \in pts(\norm{y})}{pts(o^t_i.\norm{f}) \subseteq pts($x$)}} &
		\multicolumn{2}{c}{\ruledef{o^t_i \in pts(\norm{x})}{pts(\norm{y}) \subseteq pts(o^t_i.\norm{f})}} \bigstrut \topStrut \\[1em]

		\arrayrulecolor{gray}\hline
		\multicolumn{2}{c}{\color{gblue}\text{x} = \text{y.m}($ arg_0, ,\ldots, arg_{n-1} $)} & \multicolumn{2}{c}{\rulename{P-Call}} \bigstrut \\ 

		\multicolumn{4}{c}{
			\ruledef{o^{t}_{i} \in pts(\norm{\text{y}}) \hspace{.7em} m\textprime=dispatch(o^{t}_{i}, \norm{m})}
				{
				\{ o^{t}_{i} \} \subseteq pts(m\textprime_{\tt this}) \hspace{.7em} pts(m\textprime_{\tt ret}) \subseteq pts(\norm{x}) \\				
				\forall \: 0 \leqslant k < n:
				pts(arg_k) \subseteq pts(m_{p_k})
				}
			} \bigstrut \topStrut \\[1.8em]
		
		\arrayrulecolor{gray}\hline
	\end{tabular}
	\caption{Rules for pointer analysis.}
	\label{formal:pta}
\end{figure}

In a reflection-free program, only
five types of statements exist.
In \rulename{P-New}, $o_i^t$ uniquely identifies 
the abstract object created as an instance of $t$ at 
this allocation site, labeled by \emph{i}. 
In \rulename{P-Copy}, the points-to facts flow from
the RHS to the LHS of a copy statement.
In \rulename{P-Load} and \rulename{P-Store}, the
fields of an abstract object $o_i^t$ are distinguished. 

In \rulename{P-Call} (for non-reflective calls), 
the function\linebreak $dispatch(o_i^t,m)$ 
is used to resolve the virtual dispatch of method $m$ on
the receiver object $o_i^t$ to be $m'$.
We assume that $m'$ has a 
formal parameter $m_{this}'$ for the receiver object and
$m_{p_0}',\dots,m_{p_{n-1}}'$ for the remaining parameters, 
and a pseudo-variable $m_{ret}'$ is used to hold the return 
value of~$m'$.

\subsection{Reflection Analysis: Cases 1 and 2}

\Cref{formal:infer} gives the rules for resolving
reflective calls for Cases~1 and~2 
in Figure~\ref{method:overview} simultaneously. \ripple
is the first to combine both cases in reflection analysis for
Android apps. In Case~1,
regular string inference for constant class and method
names is applied. In Case~2, type inference for non-constant but non-null
class and method names is applied. As discussed
earlier, reflective field accesses are omitted. 
Recall that $\isubtype{}{}$ denotes the subtyping relation.

Let us now examine the five rules for Cases 1 and 2. 
\rulename{C12-ForName} handles a \texttt{Class.forName(cName)} call.
For the auxiliary function
$toClass:   \mathbb{O} \to \mathbb{C}$,
$toClass(o_\_^{\tt String})$ takes a string object $o_\_^{\tt String}$ and 
returns its corresponding class metaobject. If 
$o_\_^{\tt String}$ is a constant, then $toClass(o)=c^t$, where 
$t$ is the class type named by \texttt{cName}. 
Otherwise, \linebreak $toClass(o_\_^{\tt String})=c^{u}$, since \texttt{cName}
is a non-constant but no-null string. 
In this rule, \texttt{clz} is thus made to point to 
either $c^t$ or $c^u$ accordingly. In the case of $c^u$, the
missing type may be inferred when 
it is used in a subsequent reflective call. 

\rulename{C12-GetMtd} handles a
\texttt{clz.getMethod(mName,\ldots)} \linebreak call analogously.
For the auxiliary function,
$toMtdSig: \mathbb{C} \times \mathbb{O} \to \mathcal{P}(\mathbb{S})$, 
$toMtdSig(c^-, o^{\tt String}_\_)$ returns the set of method signatures
for the methods declared in or inherited by the class 
$c^-$ with their method name identified by \texttt{mName}. 
If $c^-=c^u$ but \texttt{mName}
is a string constant, say, 
``\texttt{foo}'', then \texttt{foo} is recorded: 
$toMtdSig(c^-, o_\_^{\tt String})=\{(u,\texttt{foo},u)\}$.
If \texttt{mName} is a non-constant but no-null string, 
then $toMtdSig(c^-, o_\_^{\tt String})\linebreak=\{(u)\}$.
Therefore, this rule distinguishes two cases for every signature $s
\in toMtdSig(c^-, o^{\tt String}_\_)$.  
If $c^-=c^t$ is a statically known type $t$,
a method metaobject $m^t_s$ is created.
Otherwise, a method metaobject $m^u_s$ is created. In both cases,
the missing information in a method metaobject may be inferred
when it is used in a subsequent reflective call.

\rulename{C12-New} handles reflective object allocation at a call to
\texttt{x = (T) clz.newInstance()}, where \texttt{T}
symbolizes an
intra-procedurally post-dominating type cast for the call if it
exists or \texttt{java.lang.Object} otherwise. 
If $c^-=c^t$ is a statically known type $t$, then 
\texttt{x = clz.newInstance()} degenerates into
\texttt{x = new t()}  and can thus be handled as in \rulename{P-New}.
Otherwise, $c^-=c^u$. If $\texttt{T} \neq \texttt{java.lang.Object}$,
then $u$ is inferred to be \texttt{T} or any of its subtypes. 

\begin{figure}[t]
	\vspace*{1ex}
	\centering
	\begin{adjustbox}{width=\linewidth}
		\begin{tabular}{l@{\hspace{1cm}}l}

		\arrayrulecolor{gray}\hline
		\color{gblue}\text{Class} \norm{clz} = \text{Class}.forName(\text{cName}) & \rulename{C12-ForName} \bigstrut \\

		\multicolumn{2}{c}{
			\ruledef{o^{\tt String}_{\_} \in pts(\norm{cName}) \hspace{.7em} { c}^{\_} = toClass(o^{\tt String}_{\_})}
				{
					pts(\norm{clz}) \supseteq 
					\left 
					\{ \hspace{-0.7em}
						\begin{array}{@{\hspace{2ex}}l@{\hspace{2ex}}l}
							\{ { c}^t \} & {\tt if} \; { c}^{\_} = { c}^t \\
							\{ { c}^{ u} \} & {\tt if} \; { c}^{\_} = { c}^{ u}
						\end{array}
					\right. 
				}
		} \bigstrut \topStrut \\[2.2em]

		\arrayrulecolor{gray}\hline
		\color{gblue}\text{Method} \norm{mtd} = \norm{clz}\text{.getMethod(mName}, \_) & \rulename{C12-GetMtd} \bigstrut \\
		
		\multicolumn{2}{c}{
			\ruledef{o^{\tt String}_{\_} \in pts(\norm{mName}) \hspace{.7em} { c}^{\_} \in pts(\norm{clz}) \hspace{.7em} s \in toMtdSig({ c}^{\_}, o^{{\tt String}}_{\_})}
				{
pts(\norm{mtd}) \supseteq 
\left
\{ \hspace{-0.7em}
\begin{array}{@{\hspace{2ex}}l@{\hspace{2ex}}l}
\{ { m}^t_s \} & {\tt if} \; { c}^{\_} = { c}^t \\
\{ { m}^{ u}_s \} & {\tt if} \; { c}^{\_} = { c}^{ u}  \\
\end{array}
\right.
}
} \bigstrut \topStrut \\[2.em]

		\arrayrulecolor{gray}\hline
		$i:$ \color{gblue}\norm{x} = (\norm{T}) \norm{clz}.\norm{newInstance}() & \rulename{C12-New} \bigstrut \\

		\multicolumn{2}{c}{
			\ruledef{{\tt c}^{\_} \in pts(\norm{clz}) \hspace{.7em} }
			{
				pts(\norm{x}) \supseteq 
				\left
				\{ \hspace{-0.7em}
					\begin{array}{@{\hspace{2ex}}l@{\hspace{2ex}}l}
						\{ o^{t}_{i} \} & {\tt if} \; {c}^{\_} = {c}^{t} \\
						\{ o^{t}_{i} \mid\isubtype{t}{T} \} & {\tt if} \; {c}^{\_} = {c}^{u}
					\end{array}
				\right.
			}
		} \bigstrut \topStrut \\[2.2em]

		\arrayrulecolor{gray}\hline
		\color{gblue}\_ = \text{mtd}.\text{invoke}(\text{y}, \_) & \rulename{C12-InvType} \bigstrut \\ 

		\multicolumn{2}{c}{
			\ruledef{o^{t}_{\_} \in pts(\text{y}) \quad { m}^{ u}_{\_} \in pts(\norm{mtd})}
{ pts(\norm{mtd}) \supseteq \{ {\tt m}^{t}_{\_} \} }
		} \bigstrut \topStrut \\[0.8em]

		\arrayrulecolor{gray}\hline
		\color{gblue}\text{\_} = \text{(T)} \text{mtd}.\text{invoke}(\_, args) & \rulename{C12-InvSig} \bigstrut \\
		
		\multicolumn{2}{c}{
			\ruledef{
				\begin{array}{@{\hspace{2ex}}c@{\hspace{2ex}}c}
					{{m}^-_{u} \in pts(\norm{mtd}) \hspace{.7em} \ifamily{t}{T} \hspace{.7em} p \in toParasTys(args)}
				\end{array}
			}
			{
				pts(\norm{mtd}) \supseteq
				\{ \, {m}^-_s \; | \; s.{\tt para} = p \wedge s.{\tt ret} = t \, \} 
			}
		} \bigstrut \rule{0pt}{2em} \\[1em]
		\arrayrulecolor{gray}\hline
	\end{tabular}
	\end{adjustbox}
	\caption{Rules for Cases 1 and 2 in Figure~\ref{method:overview}.
	}
	\label{formal:infer}
\end{figure}

There are two rules for handling reflective method invocation.
To infer a target method invoked reflectively, we need to infer
its class type, which is handled by 
\rulename{C12-InvType}, and its signature, which is handled by
\rulename{C12-InvSig}.

\rulename{C12-InvType} is simple.
The class type of a method metaobject
${m}^{u}_{\_}$ is inferred to be $m^t_{\_}$ for 
every possible dynamic type $t$ of 
every receiver object pointed to by $y$.

\rulename{C12-InvSig} is slightly more involved in handling
\texttt{(T) mtd.invoke($\_$, args)}, where \texttt{T} is
defined identically as in \rulename{C12-New}. 
This rule attempts
to infer the missing information in the signature $s$ of a method
from its \texttt{args} and its possible return type. Here, we write
$s.{\tt para}$ and $s.{\tt ret}$ to identity its parameter types 
and return type, respectively. The typing
relation $\ifamily{}{}$ is defined by distinguishing two cases. First,
$\ifamily{u}{\texttt{java.lang.Object}}$. Second, if
$t$ is not \texttt{java.lang.Object},  then
$\ifamily{t'}{t}$ holds if and only if 
$\isubtype{t'}{t}$ or $\isubtype{t}{t'}$ holds. Therefore,
$s.{\tt ret}$ is deduced from a post-dominating cast \texttt{T}
(which is not
\texttt{java.lang.Object}). As for $s.{\tt para}$, we infer it
intra-procedurally from \texttt{args}. For the auxiliary function
$ toParaTys: \bigcup^{\infty}_{i=0} \mathbb{V}^i 
\to \mathcal{P}(\mathbb{S})$,
\linebreak
$toParaTys(\texttt{args})$ returns the set of 
parameter types of a target method invoked with its argument list \texttt{args},
computed only intra-procedurally for efficiency reasons.
If \texttt{args} is not defined locally, i.e., not
in the same method containing the reflective method call,
then
$toParaTys(\texttt{args})=\varnothing$. Otherwise,
let $D_i$ be the set of declared types of all possible variables
assigned to the $i$-th argument $args[i]$.
Let $P_i = \{ t'  \mid  t \in D_i \wedge 
(\isubtype{t'}{t} \vee \isubtype{t}{t'}) \}$.
Then, $toParaTys(\texttt{args})=P_0\times \cdots \times P_{n-1}$.
With $s.{\tt para}$
or $s.{\tt ret}$ inferred for $m^-_u$, \texttt{mtd}  is 
made to point to a new method metaobject $m^-_s$, where $s$ contains
the missing information in $u$ deduced via inference.

\subsection{Reflection Analysis: Case 3}

\Cref{formal:iae} gives the rules for resolving reflective
calls for Case~3 in Figure~\ref{method:overview}. There are 
four rules for handling three categories of incomplete information,
\textsc{Null-Name}, 
\textsc{Missing-RecvObj} and 
\textsc{Missing-RetObj}, as discussed in Section~\ref{sec:meth}.

\begin{figure}[ht]
	\centering
	\begin{adjustbox}{width=\linewidth}
		\begin{tabular}{ll}
		\arrayrulecolor{gray}\hline
		\color{gblue}\text{Class clz} = \text{Class}.forName(\text{cName}) & \rulename{C3-ForName}  \bigstrut \\

		\multicolumn{2}{c}{
			\ruledef{pts(\norm{cName}) = \varnothing}
{ pts(\norm{clz}) \supseteq \{ {c}^{u} \} }
		} \bigstrut \topStrut \\[1em]

		\arrayrulecolor{gray}\hline
		\color{gblue} \text{Method mtd = clz.getMethod(mName, \_)} & \rulename{C3-GetMtd} \bigstrut \\

		\multicolumn{2}{c}{
			\ruledef{pts(\text{mName}) = \varnothing \quad { c}^{\_} \in pts(\norm{clz})}
				{
					pts(\norm{mtd}) \supseteq
					\left
					\{ \hspace{-0.7em}
					\begin{array}{@{\hspace{2ex}}l@{\hspace{2ex}}l}
						\{ { m}^t_{ u} \} & {\tt if} \; {c}^{\_} = { c}^t \\
						\{ { m}^{ u}_{ u}\} & {\tt if} \; {c}^{\_} = {c}^{ u}
					\end{array}
					\right.
				}
		} \bigstrut \topStrut \\[2.2em]

		\arrayrulecolor{gray}\hline
		$i:$ \color{gblue} \_ = \text{mtd.invoke(y, \_)} &  \rulename{C3-InvRecv} \bigstrut \\
		\multicolumn{2}{c}{
			\ruledef{
				t'' \in ( \{ \, t \; | \; {\tt m}^{t}_{\_} \in pts(\norm{mtd}) \, \} \setminus \{ \, t'  \mid  o_\_^{t} \in pts(\norm{y}) \wedge t<: t' \, \} ) \\ t'' \neq \texttt{java.lang.Object}
				}
{
	pts(\norm{y}) \supseteq \{ o^{t'''}_{i} \mid 
	\ifamily{t'''}{t''}
\} }
				} \bigstrut \topStrut \\[1.2em]

		\arrayrulecolor{gray}\hline
		$i:$ \color{gblue} \text{x = mtd.invoke(\_, \_)} & \rulename{C3-InvRet} \bigstrut \\

		\multicolumn{2}{c}{
			\ruledef{ 
{m}^-_s \in pts(\norm{mtd}) \quad  s.{\tt ret} =t 
\quad
\isubtype{t'}{t} \quad  \forall\ o_\_^{t''} \in \pts(x) :
t'' \not\!<: t \\
t \neq \texttt{java.lang.Object} 
}
{ pts(\norm{x}) \supseteq \{ o^{t'} \} }
		} \bigstrut \topStrut \\[1.2em]
			
		\arrayrulecolor{gray}\hline

		\end{tabular}
		\end{adjustbox}
	\caption{Rules for Case 3 in Figure~\ref{method:overview}.
	\label{formal:iae}}
\end{figure}

\rulename{C3-ForName} and \rulename{C3-GetMtd} deal with
\textsc{Null-Name}
by treating null as a non-constant string and then
resorting to 
\rulename{C12-New}, \rulename{C12-InvType} and
\rulename{C12-InvSig} to
infer the missing information. 
\rulename{C3-ForName} handles a \texttt{Class.\linebreak forName(cName)} call 
with \texttt{cName = null} identically as how 
\rulename{C12-ForName} 
handles a \texttt{Class.forName(cName)} call 
when \texttt{cName} is a non-constant string. Similarly,
\rulename{C3-GetMtd} handles a \texttt{clz.getMethod(mName,\ldots)}
call with \texttt{mName = null} identically as  how
\rulename{C12-GetMtd} 
handles a\linebreak \texttt{clz.getMethod(mName,\ldots)} call 
when \texttt{mName} is a non-constant string. 

\rulename{C3-InvRecv} handles \textsc{Missing-RecvObj} by inferring
the missing receiver objects pointed to by \texttt{y} from the known
class types of all possibly invokable target methods, except that
\texttt{java.lang.Object} is excluded for precision reasons. This rule
covers an important special case when $\pts(y)=\varnothing$.

\rulename{C3-InvRet} handles \textsc{Missing-RetObj} by inferring
the missing objects returned from a target method that is unmodeled
(with its body missing) but available for analysis from the
return type $s.{\tt ret}$ of its signature $s$.
Objects of all possible subtypes of $s.{\tt ret}$ are
included in $\pts(x)$,  unless $x$ already points to
an object of one of these subtypes.

\subsection{Transforming Reflective to Regualar Calls}

\newcommand{\n}[1]{\textnormal{#1}}

Fig.~\ref{fig:trans} shows how to
transform a reflective into a regular call,
which will be analyzed by pointer analysis.

\begin{figure}[ht]
	\centering
	\begin{adjustbox}{width=\linewidth}
		\begin{tabular}{l@{\hspace{3cm}}l}
		\arrayrulecolor{gray}\hline
		\color{gblue}\text{x} = \text{mtd}.invoke(\text{y}, args) & \rulename{T-Inv}  \bigstrut \\

		\multicolumn{2}{c}{
			\ruledef{
\texttt{m}^t_{s} \in pts\n{(mtd)} \quad m' \in MTD(\texttt{m}^t_{s}) \quad o^\phd_i \in \pts\n{(args)} \\
 o^{t'}_j \in \pts(o^\phd_i.arr)\quad t''\ \mbox{is declaring type
 of}\ m_{p_k}'\quad k\in[0,n-1] \quad \isubtype{t'}{t''} \\
			}
{ 
	\{o^{t'}_j\} \subseteq \pts(\n{arg}_k) \quad
	\n{x} = \n{y}.m'(\n{arg}_0,...,\n{arg}_{n-1})
}
		} \bigstrut \topStrut \\[1em]
		\arrayrulecolor{gray}\hline
		\end{tabular}
		\end{adjustbox}
\caption{Rule for \emph{Transformation}.
\label{fig:trans}}
\end{figure}

For the auxiliary function
$MTD: \mathbb{M} \rightarrow \mathcal{P}({M})$,
where $M$ is the set of methods in the program,
$MTD(m_s^t)$ is the 
standard lookup function for finding the methods in $M$
according
to a declaring class $t$ and a signature $s$ for a 
method metaobject, except that
(1) the return type in $s$
is also considered and (2) any $u$ in $s$
is treated as a wild card.

As discussed earlier, 
\texttt{args} points to a 1-D array of type \texttt{Object[]},
with its elements collapsed to a single field $arr$ during the
pointer analysis.
Let \texttt{arg$_0$},\ldots, \texttt{arg$_{n-1}$} be the $n$
freshly created arguments to be passed to each potential
target method $m'$ found by in $MTD(m_s^t)$.
Let $m_{p_0}',\dots,m_{p_{n-1}}'$ be the $n$ parameters
(excluding $this$) of $m'$, such that the declaring type of $m_{p_k}'$
is $t''$. We include $o_j^{t'}$ to $\pts({\tt arg}_k)$ only
when $\isubtype{t'}{t''}$ holds in order to filter out the 
objects that cannot be assigned to $m_{p_k}'$. Finally,
the regular call obtained can be analyzed
by \rulename{P-Call} in Figure~\ref{formal:pta}.

\subsection{Examples}
\label{sec:ex}

Let us apply \ripple to the app in Figure~\ref{motivating:intent}.
Due to an undetermined intent in line 3,
\texttt{cName} is null in line~4. By applying
\rulename{C3-ForName} to \texttt{Class.forName(cName)} in line 5, 
we obtain $\pts(\texttt{clz})=\{c^u\}$. In line 6, we apply
\rulename{C12-New} to obtain
$\pts(\texttt{mmBase})=\{
o_6^{\tt\scriptsize MMBaseActivity},
o_6^{\tt\scriptsize AdViewOverlayActivity},\linebreak
o_6^{\scriptsize\tt BridgeMMMedia\$PickerActivity},
o_6^{\scriptsize\tt CachedVideoPlayerActivity},
o_6^{\scriptsize\tt VideoPlayerActivity} 
\}$, as these five class types are deduced from 
the post-dominating cast \texttt{MMBaseActivity} for 
the call to \texttt{clz.newInstance()} in line 6. As a result,
\ripple discovers
\num{3928} caller-callee edges 
in lines 5 -- 7 (directly or indirectly), 
thereby enabling
\FlowDroid \cite{flowdroid} to detect
49 sensitive data leaks along these edges.

Let us apply \ripple to the app in Figure~\ref{motivating:framework}.
As \texttt{getDefault} is a hidden API method and thus unavailable for
analysis, \texttt{telephonyManager = null}. By applying
\rulename{C3-InvRecv} to \texttt{getSubIdMtd.invoke(telephonyManager)} in
line 6, we can deduce precisely that 
$\pts(\texttt{telephonyManager})  =\linebreak \{o_6^{\tt TelephonyManager}\}$.
In this app, \texttt{TelephonyManager} has no subtypes but only
\texttt{java.lang.Object} as its supertype.
This allows the potential leak highlighted in
Figure~\ref{motivating:framework} to be detected.
If
\texttt{getDefault} were unmodeled but available for analysis,
then we would be able to apply \rulename{C3-InvRet} to 
\texttt{getDefMtd.invoke(null)} in line 4 to infer the missing
object returned:
$\pts(\texttt{telephonyManager})  =\{o_4^{\tt TelephonyManager}\}$.
Again, the same sensitive data leak will be detected.

\section{Evaluation}
\label{section:eval}
\label{sec:eval}

We demonstrate that \ripple is significantly more effective than
string inference 
for real-world Android apps by addressing the following four research
questions:
\begin{itemize}[itemsep=0ex,leftmargin=*]
    \item{\bf{RQ1.}} Is \ripple capable of discovering more 
reflective targets, i.e., more sound
	    than string inference?
    \item{\bf{RQ2.}} Can \ripple achieve this with a good precision?
    \item{\bf{RQ3.}} Does \ripple scale for real-world Android apps?
    \item{\bf{RQ4.}} Is \ripple effective in enabling 
	    existing Android security analyses to
	    detect security vulnerabilities?
\end{itemize}

To answer RQ1 -- RQ3, we examine the reflective targets resolved 
by \ripple in Android apps. To answer RQ4, we investigate how \ripple
enables \FlowDroid \cite{flowdroid}, a taint analysis for Android apps,
to find more sensitive data leaks.

\paragraph*{\small\bf Android Apps}

Real-world Android apps rather than synthetic benchmarks are used
in our experiments. We examined the top-chart
free applications from Google Play downloaded
on 15 April 2016, which are the most 
popular apps in the official app store. A set of 17 apps 
is selected, such that they
exhibit a wide range of incomplete information, with null
class and method names, as discussed
in Section~\ref{sec:iie}, and they are scalable under
\FlowDroid within 2 hours.

\begin{table*}[ht]
  \centering
  \caption{Soundness and precision. \protect\fauxsc{Ripple} always finds every
  true reflective target that \protect\fauxsc{Strinf} does. The
  numbers in \textbf{bold} indicate that \protect\fauxsc{Ripple}
  finds more
	  true targets than
	  \protect\fauxsc{Strinf}. A reflective call in app is considered to be reachable if it
	  can be reached from the harness \texttt{main()} in its call graph.
  }
    \begin{adjustbox}{width=\linewidth}
    \begin{tabular}{l|r|r|r|c|r|r|r|c||r|r|r|c|r|r|r|c}
    \hline
    \multicolumn{1}{c|}{\multirow{4}[8]{*}{\textbf{App (Package Name)}}} & \multicolumn{8}{c||}{{\strinf}} & \multicolumn{8}{c}{{\ripple}} \bigstrut\\
    \cline{2-17}
    \multicolumn{1}{c|}{} & \multicolumn{4}{c|}{\textbf{Class.newInstance()}} & \multicolumn{4}{c||}{\textbf{Method.invoke()}} & \multicolumn{4}{c|}{\textbf{Class.newInstance}} & \multicolumn{4}{c}{\textbf{Method.invoke()}} \bigstrut\\
    \cline{2-17}
    \multicolumn{1}{c|}{} & \multicolumn{2}{c|}{\textbf{\#Calls}} & \multicolumn{2}{c|}{\textbf{\#Targets}} & \multicolumn{2}{c|}{\textbf{\#Calls}} & \multicolumn{2}{c||}{\textbf{\#Targets}} & \multicolumn{2}{c|}{\textbf{\#Calls}} & \multicolumn{2}{c|}{\textbf{\#Targets}} & \multicolumn{2}{c|}{\textbf{\#Calls}} & \multicolumn{2}{c}{\textbf{\#Targets}} \bigstrut\\
    \cline{2-17}
    \multicolumn{1}{c|}{} & \multicolumn{1}{c|}{\textbf{Reachable}} & \multicolumn{1}{c|}{\textbf{Resolved}} & \multicolumn{1}{c|}{\textbf{Resolved}} & \multicolumn{1}{c|}{\textbf{True}} & \multicolumn{1}{c|}{\textbf{Reachable }} & \multicolumn{1}{c|}{\textbf{Resolved }} & \multicolumn{1}{c|}{\textbf{Resolved }} & \multicolumn{1}{c||}{\textbf{True}} & \multicolumn{1}{c|}{\textbf{Reachable }} & \multicolumn{1}{c|}{\textbf{Resolved}} & \multicolumn{1}{c|}{\textbf{Resolved }} & \multicolumn{1}{c|}{\textbf{True}} & \multicolumn{1}{c|}{\textbf{Reachable }} & \multicolumn{1}{c|}{\textbf{Resolved }} & \multicolumn{1}{c|}{\textbf{Resolved}} & \multicolumn{1}{c}{\textbf{True}} \bigstrut\\
    \hline
    \multicolumn{1}{l|}{com.facebook.orca} & \multicolumn{1}{c}{0} & \multicolumn{1}{c}{0} & \multicolumn{1}{c}{0} & 0     & \multicolumn{1}{c}{7} & \multicolumn{1}{c}{0} & \multicolumn{1}{c}{0} & 0     & \multicolumn{1}{c}{0} & \multicolumn{1}{c}{0} & \multicolumn{1}{c}{0} & 0     & \multicolumn{1}{c}{7} & \multicolumn{1}{c}{1} & \multicolumn{1}{c}{7} & \textbf{7} \bigstrut[t]\\

    \multicolumn{1}{l|}{com.netmarble.sknightsgb} & \multicolumn{1}{c}{2} & \multicolumn{1}{c}{0} & \multicolumn{1}{c}{0} & 0     & \multicolumn{1}{c}{11} & \multicolumn{1}{c}{4} & \multicolumn{1}{c}{8} & 4     & \multicolumn{1}{c}{2} & \multicolumn{1}{c}{2} & \multicolumn{1}{c}{26} & \textbf{11} & \multicolumn{1}{c}{11} & \multicolumn{1}{c}{4} & \multicolumn{1}{c}{8} & 4 \\

    \multicolumn{1}{l|}{com.productmadness.hovmobile} & \multicolumn{1}{c}{1} & \multicolumn{1}{c}{0} & \multicolumn{1}{c}{0} & 0     & \multicolumn{1}{c}{7} & \multicolumn{1}{c}{5} & \multicolumn{1}{c}{31} & 29    & \multicolumn{1}{c}{1} & \multicolumn{1}{c}{1} & \multicolumn{1}{c}{11} & \textbf{1} & \multicolumn{1}{c}{7} & \multicolumn{1}{c}{5} & \multicolumn{1}{c}{31} & 29 \\

    \multicolumn{1}{l|}{com.facebook.moments} & \multicolumn{1}{c}{0} & \multicolumn{1}{c}{0} & \multicolumn{1}{c}{0} & 0     & \multicolumn{1}{c}{5} & \multicolumn{1}{c}{0} & \multicolumn{1}{c}{0} & 0     & \multicolumn{1}{c}{0} & \multicolumn{1}{c}{0} & \multicolumn{1}{c}{0} & 0     & \multicolumn{1}{c}{5} & \multicolumn{1}{c}{1} & \multicolumn{1}{c}{7} & \textbf{7} \\

    \multicolumn{1}{l|}{me.msqrd.android} & \multicolumn{1}{c}{0} & \multicolumn{1}{c}{0} & \multicolumn{1}{c}{0} & 0     & \multicolumn{1}{c}{11} & \multicolumn{1}{c}{4} & \multicolumn{1}{c}{4} & 4     & \multicolumn{1}{c}{0} & \multicolumn{1}{c}{0} & \multicolumn{1}{c}{0} & 0     & \multicolumn{1}{c}{11} & \multicolumn{1}{c}{4} & \multicolumn{1}{c}{4} & 4 \\

    \multicolumn{1}{l|}{com.nordcurrent.canteenhd} & \multicolumn{1}{c}{3} & \multicolumn{1}{c}{0} & \multicolumn{1}{c}{0} & 0     & \multicolumn{1}{c}{16} & \multicolumn{1}{c}{5} & \multicolumn{1}{c}{6} & 5     & \multicolumn{1}{c}{4} & \multicolumn{1}{c}{3} & \multicolumn{1}{c}{27} & \textbf{10} & \multicolumn{1}{c}{20} & \multicolumn{1}{c}{5} & \multicolumn{1}{c}{6} & 5 \\

    \multicolumn{1}{l|}{com.ea.game.simcitymobile\_row} & \multicolumn{1}{c}{0} & \multicolumn{1}{c}{0} & \multicolumn{1}{c}{0} & 0     & \multicolumn{1}{c}{6} & \multicolumn{1}{c}{2} & \multicolumn{1}{c}{2} & 2     & \multicolumn{1}{c}{0} & \multicolumn{1}{c}{0} & \multicolumn{1}{c}{0} & 0     & \multicolumn{1}{c}{6} & \multicolumn{1}{c}{2} & \multicolumn{1}{c}{2} & 2 \\

    \multicolumn{1}{l|}{com.imangi.templerun} & \multicolumn{1}{c}{0} & \multicolumn{1}{c}{0} & \multicolumn{1}{c}{0} & 0     & \multicolumn{1}{c}{7} & \multicolumn{1}{c}{1} & \multicolumn{1}{c}{1} & 1     & \multicolumn{1}{c}{0} & \multicolumn{1}{c}{0} & \multicolumn{1}{c}{0} & 0     & \multicolumn{1}{c}{7} & \multicolumn{1}{c}{1} & \multicolumn{1}{c}{1} & 1 \\

    \multicolumn{1}{l|}{com.rovio.angrybirds} & \multicolumn{1}{c}{3} & \multicolumn{1}{c}{1} & \multicolumn{1}{c}{1} & 1     & \multicolumn{1}{c}{10} & \multicolumn{1}{c}{3} & \multicolumn{1}{c}{3} & 3     & \multicolumn{1}{c}{3} & \multicolumn{1}{c}{2} & \multicolumn{1}{c}{6} & \textbf{6} & \multicolumn{1}{c}{14} & \multicolumn{1}{c}{8} & \multicolumn{1}{c}{13} & \textbf{11} \\

    \multicolumn{1}{l|}{com.sgn.pandapop.gp} & \multicolumn{1}{c}{0} & \multicolumn{1}{c}{0} & \multicolumn{1}{c}{0} & 0     & \multicolumn{1}{c}{13} & \multicolumn{1}{c}{3} & \multicolumn{1}{c}{3} & 3     & \multicolumn{1}{c}{0} & \multicolumn{1}{c}{0} & \multicolumn{1}{c}{0} & 0     & \multicolumn{1}{c}{13} & \multicolumn{1}{c}{3} & \multicolumn{1}{c}{3} & 3 \\

    \multicolumn{1}{l|}{\leftspecialcell{com.gameloft.android. \\ ANMP.GloftA8HM}} & \multicolumn{1}{c}{1} & \multicolumn{1}{c}{1} & \multicolumn{1}{c}{16} & 16    & \multicolumn{1}{c}{9} & \multicolumn{1}{c}{0} & \multicolumn{1}{c}{0} & 0     & \multicolumn{1}{c}{1} & \multicolumn{1}{c}{1} & \multicolumn{1}{c}{16} & 16    & \multicolumn{1}{c}{9} & \multicolumn{1}{c}{0} & \multicolumn{1}{c}{0} & 0 \\

    \multicolumn{1}{l|}{com.appsorama.kleptocats} & \multicolumn{1}{c}{2} & \multicolumn{1}{c}{2} & \multicolumn{1}{c}{5} & 5     & \multicolumn{1}{c}{3} & \multicolumn{1}{c}{0} & \multicolumn{1}{c}{0} & 0     & \multicolumn{1}{c}{2} & \multicolumn{1}{c}{2} & \multicolumn{1}{c}{5} & 5     & \multicolumn{1}{c}{3} & \multicolumn{1}{c}{1} & \multicolumn{1}{c}{6} & \textbf{4} \\

    \multicolumn{1}{l|}{air.au.com.metro.DumbWaysToDie} & \multicolumn{1}{c}{1} & \multicolumn{1}{c}{0} & \multicolumn{1}{c}{0} & 0     & \multicolumn{1}{c}{18} & \multicolumn{1}{c}{8} & \multicolumn{1}{c}{34} & 34    & \multicolumn{1}{c}{1} & \multicolumn{1}{c}{1} & \multicolumn{1}{c}{5} & \textbf{5} & \multicolumn{1}{c}{21} & \multicolumn{1}{c}{11} & \multicolumn{1}{c}{37} & \textbf{37} \\

    \multicolumn{1}{l|}{com.ketchapp.twist} & \multicolumn{1}{c}{3} & \multicolumn{1}{c}{2} & \multicolumn{1}{c}{5} & 5     & \multicolumn{1}{c}{8} & \multicolumn{1}{c}{2} & \multicolumn{1}{c}{2} & 2     & \multicolumn{1}{c}{3} & \multicolumn{1}{c}{2} & \multicolumn{1}{c}{5} & 5     & \multicolumn{1}{c}{8} & \multicolumn{1}{c}{3} & \multicolumn{1}{c}{8} & \textbf{6} \\

    \multicolumn{1}{l|}{com.stupeflix.legend} & \multicolumn{1}{c}{4} & \multicolumn{1}{c}{2} & \multicolumn{1}{c}{2} & 2     & \multicolumn{1}{c}{1} & \multicolumn{1}{c}{0} & \multicolumn{1}{c}{0} & 0     & \multicolumn{1}{c}{4} & \multicolumn{1}{c}{3} & \multicolumn{1}{c}{13} & \textbf{5} & \multicolumn{1}{c}{1} & \multicolumn{1}{c}{0} & \multicolumn{1}{c}{0} & 0 \\

    \multicolumn{1}{l|}{com.maxgames.stickwarlegacy} & \multicolumn{1}{c}{1} & \multicolumn{1}{c}{0} & \multicolumn{1}{c}{0} & 0     & \multicolumn{1}{c}{2} & \multicolumn{1}{c}{1} & \multicolumn{1}{c}{1} & 1     & \multicolumn{1}{c}{1} & \multicolumn{1}{c}{0} & \multicolumn{1}{c}{0} & 0     & \multicolumn{1}{c}{2} & \multicolumn{1}{c}{2} & \multicolumn{1}{c}{7} & \textbf{5} \\

    \multicolumn{1}{l|}{\leftspecialcell{air.com.tutotoons.app. \\ animalhairsalon2jungle.free}} & \multicolumn{1}{c}{0} & \multicolumn{1}{c}{0} & \multicolumn{1}{c}{0} & 0     & \multicolumn{1}{c}{14} & \multicolumn{1}{c}{7} & \multicolumn{1}{c}{43} & 43    & \multicolumn{1}{c}{0} & \multicolumn{1}{c}{0} & \multicolumn{1}{c}{0} & 0     & \multicolumn{1}{c}{14} & \multicolumn{1}{c}{7} & \multicolumn{1}{c}{43} & 43 \bigstrut[b]\\

    \hline

    \multicolumn{1}{c|}{\textbf{Total}} & \multicolumn{1}{c}{\textbf{21}} & \multicolumn{1}{c}{\textbf{8}} & \multicolumn{1}{c}{\textbf{29}} & \textbf{29} & \multicolumn{1}{c}{\textbf{148}} & \multicolumn{1}{c}{\textbf{45}} & \multicolumn{1}{c}{\textbf{138}} & \textbf{131} & \multicolumn{1}{c}{\textbf{22}} & \multicolumn{1}{c}{\textbf{17}} & \multicolumn{1}{c}{\textbf{114}} & \textbf{64} & \multicolumn{1}{c}{\textbf{159}} & \multicolumn{1}{c}{\textbf{58}} & \multicolumn{1}{c}{\textbf{183}} & \textbf{168} \bigstrut\\
    \hline

    \end{tabular}
    \end{adjustbox}
  \label{eval:rq1}
  \vspace*{-2ex}
\end{table*}

\paragraph*{\small\bf State-of-the-Art Reflection Analysis}

To resolve reflection for Android apps, there are only two 
existing static techniques, with both performing
string inference (for constant strings), \checker 
\cite{sparta,checker}
and \droidRA \cite{droidra}. We cannot compare with \checker 
since its reflection analysis relies on user annotations. We cannot
compare with \droidRA either, since its latest open-source 
tool (released on 9 September 2016) is unstable. For the 17 apps selected,
\droidRA resolves some reflective targets for 6 apps, fails to produce
any outputs for another 6 apps, and crashes for the remaining 5 apps. 
These problems were reported to and confirmed by the 
authors, but they still remain despite a recent release.

Both \checker and \droidRA perform essentially regular string inference,
which is subsumed by \ripple as shown in Case~1 (for
constant strings) in Figure~\ref{method:overview}. 
\checker handles non-constant strings with
user annotations only.
Instead of comparing with \checker and \droidRA directly, we compare
\ripple with \strinf, which is a simplified \ripple that performs only
regular string inference in Case 1.

\paragraph*{\small\bf Implementation}

We have implemented \ripple in \soot~\cite{soot}, a static analysis framework 
for Android and Java programs. \ripple works with its 
\spark, a flow- and context-insensitive pointer analysis, to resolve
reflection and points-to information in a program. Based on the results
of this joint analysis, the call graph of the program, on which
many security analyses such as \FlowDroid operate, can be constructed.

Currently, \ripple is implemented in Java with about \num{3500} LOC,
handling the most significant reflective calls that
affect the static analysis of Android apps:
\texttt{Class.forName()}, 
\texttt{Class.newInstance()}, 
\texttt{Method.invoke()}, and 
all four\linebreak method-introspecting calls,  
\texttt{Method.getMethod()},
\texttt{Method.\linebreak getDeclaredMethod()}, 
\texttt{Method.getMethods()}, 
and\linebreak \texttt{Method.getDeclaredMethods()}.

\paragraph*{\small\bf Computing Platform}

Our experiments are carried out on a Xeon E5-2650 2GHz machine with 64GB RAM running Ubuntu 14.04 LTS. The time measured for analyzing an app by
a particular analysis is the average of 20 runs.

\subsection{RQ1: More Soundness}
\label{sec:RQ1}

Table~\ref{eval:rq1} compares \ripple and \strinf in terms of 
the number of reflective targets discovered at all reflective calls
to \texttt{Class.newInstance()} and \texttt{Method.invoke()}. 
For the former, a target means a reflectively created object. For the
latter, a target means a reflectively invoked method.
Each app is uniquely identified by its package name. For each of
these two methods, only its calls reachable from the harness \texttt{main()}
used during the analysis
are included. We determine whether a target is true or not by 
manual code inspection.

By design, \ripple always finds every true target that \strinf does.
In 11 out of the 17 apps,
\ripple has successfully discovered more true targets 
than \strinf. 
This highlights the importance of making
reflection analysis fully IIE-aware for Android apps,
by handling not only Case 2 as for
Java programs
\cite{elf, solar, yannisreflection, livshits} but 
also Case 3.

In the 17 Android apps, \ripple finds a total of 64
and 168 but \strinf finds only  a total of 29 and 131
true targets for \texttt{Class.newInstance()} and \texttt{Method.invoke()},
respectively. Therefore, for both methods combined, \ripple finds
232 but \strinf finds only 160 true targets in total, yielding
a net gain of 72 true targets and thus
a 45\% increase in soundness on reflection analysis.

Let us revisit some examples in \Cref{sec:motivating}.  Consider the
app in \Cref{motivating:intent}. For the call to
\texttt{clz.newInstance()} in line~6,
\ripple infers five reflectively created objects, which are all
true targets configured to provide different forms of advertisement. 
A similar pattern appears also in the app named
\textit{Dumb Ways to Die}. 
Consider now the app in \Cref{motivating:bundle}.
For the call to \texttt{clz.newInstance()} in line~4,
\ripple infers 8 reflectively created objects, which 
are all true targets used for handling eight different types of
media according to user inputs. All these targets are missed by 
\strinf, which relies only on a simple string analysis 
for string constants.

\begin{table*}[ht]
  \centering
\caption{Efficiency and effectiveness. For efficiency, the analysis times of
\protect\fauxsc{Ripple},
\protect\fauxsc{StrInf} and
\protect\fauxsc{Spark} are compared, by using the final harness 
(iteratively)
constructed for an app by \FlowDroid. For each analysis,
the number of call graph (CG) edges discovered is also given.
For effectiveness, the number of data leaks, together with
sensitive source and sink calls, found by \FlowDroid under
\protect\fauxsc{Ripple}, 
\protect\fauxsc{StrInf} and
\protect\fauxsc{Spark} are compared. 
The numbers in \textbf{bold} indicate that
  \FlowDroid reports more leaks under
\protect\fauxsc{Ripple} than 
\protect\fauxsc{StrInf}.
  }
  \begin{adjustbox}{width=\textwidth}
    \begin{tabular}{r||c|c|c|cc||cc|c|c|c||cc|c|c|c}
    \hline
    \multicolumn{1}{c|}{\multirow{2}[4]{*}{\textbf{App Package Name}}} & \multicolumn{5}{c||}{{{\spark}}} & \multicolumn{5}{c||}{{{\strinf}}} & \multicolumn{5}{c}{{\ripple}} \bigstrut\\

    \cline{2-16}
     \multicolumn{1}{c|}{} & \multicolumn{1}{c|}{\centerspecialcell{\textbf{CG} \\ \textbf{Edges}}} & \multicolumn{1}{c|}{\centerspecialcell{\textbf{Analysis} \\ \textbf{Time (s)}}} & \multicolumn{1}{c|}{\centerspecialcell{\textbf{Source} \\ \textbf{Calls}}} & \multicolumn{1}{c|}{\centerspecialcell{\textbf{Sink} \\ \textbf{Calls}}} & \multicolumn{1}{c||}{\centerspecialcell{\textbf{Total} \\ \textbf{Leaks}}}  & \multicolumn{1}{c|}{\centerspecialcell{\textbf{CG} \\ \textbf{Edges}}} & \multicolumn{1}{c|}{\centerspecialcell{\textbf{Analysis} \\ \textbf{Time (s)}}} & \multicolumn{1}{c|}{\centerspecialcell{\textbf{Source} \\ \textbf{Calls}}} & \multicolumn{1}{c|}{\centerspecialcell{\textbf{Sink} \\ \textbf{Calls}}} & \multicolumn{1}{c||}{\centerspecialcell{\textbf{Total} \\ \textbf{Leaks}}} & \multicolumn{1}{c|}{\centerspecialcell{\textbf{CG} \\ \textbf{Edges}}} & \multicolumn{1}{c|}{\centerspecialcell{\textbf{Analysis} \\ \textbf{Time (s)}}} & \multicolumn{1}{c|}{\centerspecialcell{\textbf{Source} \\ \textbf{Calls}}} & \multicolumn{1}{c|}{\centerspecialcell{\textbf{Sink} \\ \textbf{Calls}}} & \multicolumn{1}{c}{\centerspecialcell{\textbf{Total} \\ \textbf{Leaks}}} \bigstrut \rule{0pt}{1.5em} \\

    \hline

    \multicolumn{1}{l|}{com.facebook.orca} & \multicolumn{1}{c}{5583} & \multicolumn{1}{c}{2.0} & \multicolumn{1}{c}{43} & 115   & 15    & 5598  & \multicolumn{1}{c}{2.1} & \multicolumn{1}{c}{43} & \multicolumn{1}{c}{115} & 15    & 5605  & \multicolumn{1}{c}{2.2} & \multicolumn{1}{c}{43} & \multicolumn{1}{c}{115} & 15 \bigstrut[t]\\

    \multicolumn{1}{l|}{com.netmarble.sknightsgb} & \multicolumn{1}{c}{12113} & \multicolumn{1}{c}{7.9} & \multicolumn{1}{c}{194} & 208   & 92    & 12148 & \multicolumn{1}{c}{8.0} & \multicolumn{1}{c}{194} & \multicolumn{1}{c}{208} & 96    & 12779 & \multicolumn{1}{c}{8.4} & \multicolumn{1}{c}{210} & \multicolumn{1}{c}{212} & \textbf{142} \\

    \multicolumn{1}{l|}{com.productmadness.hovmobile} & \multicolumn{1}{c}{6009} & \multicolumn{1}{c}{2.7} & \multicolumn{1}{c}{119} & 113   & 41    & 6278  & \multicolumn{1}{c}{3.1} & \multicolumn{1}{c}{119} & \multicolumn{1}{c}{113} & 44    & 6480  & \multicolumn{1}{c}{3.2} & \multicolumn{1}{c}{119} & \multicolumn{1}{c}{116} & \textbf{48} \\

    \multicolumn{1}{l|}{com.facebook.moments} & \multicolumn{1}{c}{6632} & \multicolumn{1}{c}{2.5} & \multicolumn{1}{c}{39} & 145   & 10    & 6647  & \multicolumn{1}{c}{2.6} & \multicolumn{1}{c}{39} & \multicolumn{1}{c}{145} & 10    & 6654  & \multicolumn{1}{c}{2.7} & \multicolumn{1}{c}{39} & \multicolumn{1}{c}{145} & 10 \\

    \multicolumn{1}{l|}{me.msqrd.android} & \multicolumn{1}{c}{10561} & \multicolumn{1}{c}{4.7} & \multicolumn{1}{c}{127} & 107   & 24    & 11064 & \multicolumn{1}{c}{4.8} & \multicolumn{1}{c}{127} & \multicolumn{1}{c}{107} & 25    & 11064 & \multicolumn{1}{c}{4.9} & \multicolumn{1}{c}{127} & \multicolumn{1}{c}{107} & 25 \\

    \multicolumn{1}{l|}{com.nordcurrent.canteenhd} & \multicolumn{1}{c}{16698} & \multicolumn{1}{c}{9.8} & \multicolumn{1}{c}{182} & 305   & 182   & 16759 & \multicolumn{1}{c}{10} & \multicolumn{1}{c}{182} & \multicolumn{1}{c}{305} & 184   & 21625 & \multicolumn{1}{c}{12.8} & \multicolumn{1}{c}{199} & \multicolumn{1}{c}{327} & \textbf{289} \\

    \multicolumn{1}{l|}{com.ea.game.simcitymobile\_row} & \multicolumn{1}{c}{8369} & \multicolumn{1}{c}{4.1} & \multicolumn{1}{c}{111} & 170   & 41    & 8403  & \multicolumn{1}{c}{4.2} & \multicolumn{1}{c}{111} & \multicolumn{1}{c}{172} & 50    & 8403  & \multicolumn{1}{c}{4.4} & \multicolumn{1}{c}{111} & \multicolumn{1}{c}{172} & 50 \\

    \multicolumn{1}{l|}{com.imangi.templerun} & \multicolumn{1}{c}{10584} & \multicolumn{1}{c}{2.3} & \multicolumn{1}{c}{184} & 110   & 54    & 10592 & \multicolumn{1}{c}{2.5} & \multicolumn{1}{c}{184} & \multicolumn{1}{c}{110} & 54    & 10592 & \multicolumn{1}{c}{2.6} & \multicolumn{1}{c}{184} & \multicolumn{1}{c}{110} & 54 \\

    \multicolumn{1}{l|}{com.rovio.angrybirds} & \multicolumn{1}{c}{12705} & \multicolumn{1}{c}{7.0} & \multicolumn{1}{c}{244} & 180   & 66    & 13448 & \multicolumn{1}{c}{7.4} & \multicolumn{1}{c}{245} & \multicolumn{1}{c}{206} & 66    & 17384 & \multicolumn{1}{c}{10.2} & \multicolumn{1}{c}{276} & \multicolumn{1}{c}{266} & \textbf{120} \\

    \multicolumn{1}{l|}{com.sgn.pandapop.gp} & \multicolumn{1}{c}{10557} & \multicolumn{1}{c}{5.3} & \multicolumn{1}{c}{217} & 365   & 552   & 10588 & \multicolumn{1}{c}{5.9} & \multicolumn{1}{c}{217} & \multicolumn{1}{c}{365} & 575   & 10590 & \multicolumn{1}{c}{5.9} & \multicolumn{1}{c}{217} & \multicolumn{1}{c}{365} & 575 \\

    \multicolumn{1}{l|}{\leftspecialcell{com.gameloft.android. \\ ANMP.GloftA8HM}} & \multicolumn{1}{c}{15532} & \multicolumn{1}{c}{7.0} & \multicolumn{1}{c}{274} & 181   & 126   & 16015 & \multicolumn{1}{c}{7.2} & \multicolumn{1}{c}{285} & \multicolumn{1}{c}{204} & 144   & 16016 & \multicolumn{1}{c}{6.6} & \multicolumn{1}{c}{285} & \multicolumn{1}{c}{204} & 144 \\

    \multicolumn{1}{l|}{com.appsorama.kleptocats} & \multicolumn{1}{c}{3659} & \multicolumn{1}{c}{2.7} & \multicolumn{1}{c}{101} & 106   & 39    & 3707  & \multicolumn{1}{c}{2.8} & \multicolumn{1}{c}{101} & \multicolumn{1}{c}{106} & 39    & 3714  & \multicolumn{1}{c}{3.3} & \multicolumn{1}{c}{101} & \multicolumn{1}{c}{112} & 39 \\

    \multicolumn{1}{l|}{air.au.com.metro.DumbWaysToDie} & \multicolumn{1}{c}{10059} & \multicolumn{1}{c}{4.9} & \multicolumn{1}{c}{204} & 151   & 23    & 10312 & \multicolumn{1}{c}{5.3} & \multicolumn{1}{c}{204} & \multicolumn{1}{c}{151} & 26    & 11679 & \multicolumn{1}{c}{5.9} & \multicolumn{1}{c}{228} & \multicolumn{1}{c}{167} & \textbf{103} \\

    \multicolumn{1}{l|}{com.ketchapp.twist} & \multicolumn{1}{c}{14092} & \multicolumn{1}{c}{8.7} & \multicolumn{1}{c}{210} & 179   & 93    & 14144 & \multicolumn{1}{c}{8.7} & \multicolumn{1}{c}{210} & \multicolumn{1}{c}{179} & 97    & 14151 & \multicolumn{1}{c}{8.7} & \multicolumn{1}{c}{210} & \multicolumn{1}{c}{185} & \textbf{109} \\

    \multicolumn{1}{l|}{com.stupeflix.legend} & \multicolumn{1}{c}{8803} & \multicolumn{1}{c}{3.1} & \multicolumn{1}{c}{97} & 127   & 46    & 8852  & \multicolumn{1}{c}{3.4} & \multicolumn{1}{c}{97} & \multicolumn{1}{c}{127} & 47    & 9066  & \multicolumn{1}{c}{3.5} & \multicolumn{1}{c}{97} & \multicolumn{1}{c}{127} & 47 \\

    \multicolumn{1}{l|}{com.maxgames.stickwarlegacy} & \multicolumn{1}{c}{3829} & \multicolumn{1}{c}{2.6} & \multicolumn{1}{c}{76} & 26    & 5     & 3831  & \multicolumn{1}{c}{2.5} & \multicolumn{1}{c}{76} & \multicolumn{1}{c}{26} & 5     & 3844  & \multicolumn{1}{c}{2.4} & \multicolumn{1}{c}{76} & \multicolumn{1}{c}{32} & \textbf{17} \\

    \multicolumn{1}{l|}{\leftspecialcell{air.com.tutotoons.app. \\ animalhairsalon2jungle.free}} & \multicolumn{1}{c}{11788} & \multicolumn{1}{c}{5.2} & \multicolumn{1}{c}{86} & 43    & 9     & 12075 & \multicolumn{1}{c}{5.5} & \multicolumn{1}{c}{86} & \multicolumn{1}{c}{43} & 9     & 12075 & \multicolumn{1}{c}{5.7} & \multicolumn{1}{c}{86} & \multicolumn{1}{c}{43} & 9 \bigstrut[b]\\
    
    \hline

 \multicolumn{1}{c}{{Total}} & \multicolumn{1}{c}{{167573}} & \multicolumn{1}{c}{{82.5}} & \multicolumn{1}{c}{{2508}} & {2631} & {1418} & {170461} & \multicolumn{1}{c}{{86}} & \multicolumn{1}{c}{{2520}} & \multicolumn{1}{c}{{2682}} & {1486} & {181721} & \multicolumn{1}{c}{{93.4}} & \multicolumn{1}{c}{{2608}} & \multicolumn{1}{c}{{2805}} & {1796} \bigstrut\\

    \hline
    \end{tabular}
    \end{adjustbox}
  \label{eval:rq2}
  \vspace*{-2ex}
\end{table*}

\subsection{RQ2: Precision}
\label{sec:RQ2}

Table~\ref{eval:rq1} reveals also the false positive rates for
both \ripple and \strinf. 
\ripple finds a total of 297 reflective targets
with 232 true targets, representing a false positive rate of 21.9\%.
\strinf finds a total of 167 reflective targets
with 160 true targets, representing a false positive rate of 4.2\%.

Due to 72 more true targets discovered,
as discussed in Section~\ref{sec:RQ1}, \ripple is regarded to
exhibit a satisfactory precision for Android apps. For many security 
analyses such as security vetting and malware detection, and even debugging,
it is important to analyze reflection-related program behaviors 
even if doing so may cause some false warnings to be triggered.
Consider the app in \Cref{motivating:library}. For the call to
\texttt{logMtd.invoke(null, tag, msg)} in line 6,
\ripple infers its target to be the six methods, 
\texttt{i()},
\texttt{d()},
\texttt{w()},
\texttt{e()},
\texttt{v()} and
\texttt{wtf()} in class \texttt{Log}, where the last two
are false positives, enabling 
\FlowDroid to report 12 leaks via \texttt{msg} from two 
data sources, with 4 from 
\texttt{v()} and
\texttt{wtf()} being false positives.

We can lift the  precision of \ripple by improving the precision of
its collaborating analyses. Currently, \ripple works with \spark,
a flow- and context-insensitive pointer analysis, in \soot. \ripple
will be more precise if a more precise, say, an object-sensitive pointer
analysis~\cite{Ana02} is used, instead.

In addition, \ripple will also be more precise if it works
simultaneously with other static analyses,
such as intent analysis, for Android apps.
The code snippet in \Cref{eval:imprecise} taken from \textit{Seven Knights} illustrates this proposition.

\begin{figure}[h]
    \includegraphics[width=\linewidth]{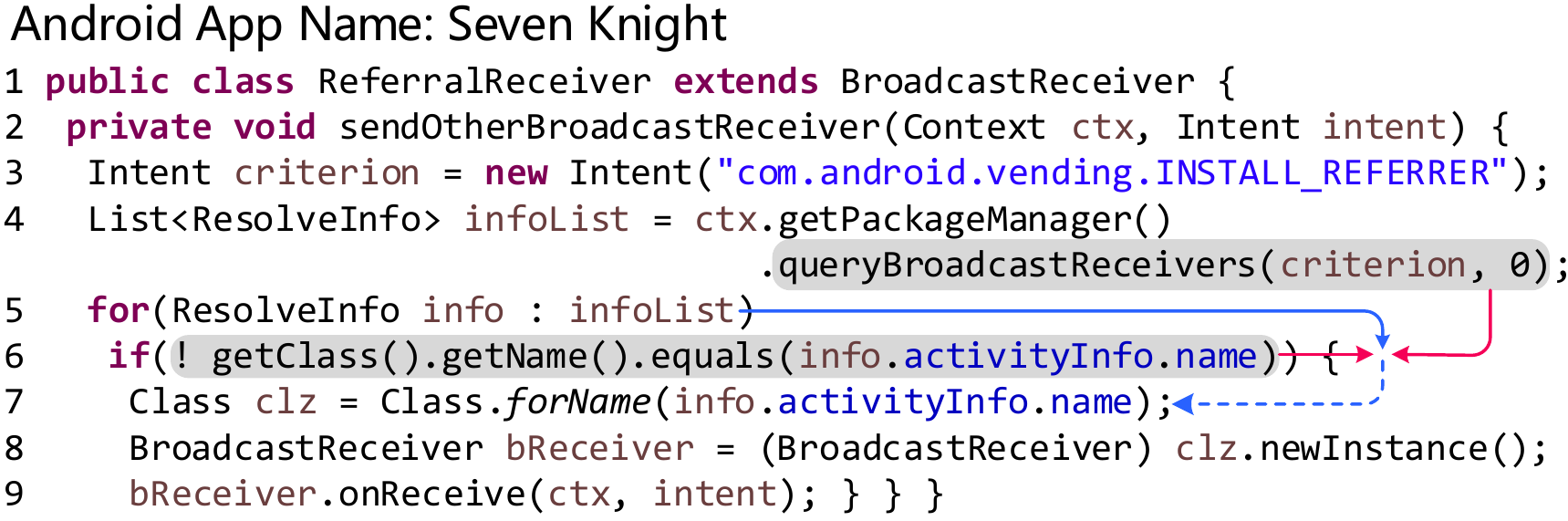}
    \caption{Improving precision in 
	    \protect\fauxsc{Ripple} by working with intent analysis. Here,
    \protect \tikz[baseline=-0.5ex] \protect \draw [line width = 0.8pt, draw = arrowblue, arrows = {-latex}, fill = arrowblue] (0, 0) -- (0.7, 0);
    denotes data-flow and
    \protect \tikz[baseline=-0.5ex] \protect \draw [densely dashed, line width = 0.8pt, draw = arrowblue, arrows = {-latex}, fill = arrowblue] (0, 0) -- (0.7, 0);
    denotes the class names discovered after considering the constraints denoted by
    \protect \tikz[baseline=-0.5ex] \protect \draw [line width = 0.8pt, draw = arrowred, arrows = {-latex}, fill = arrowred] (0, 0) -- (0.7, 0);.
    }
    \label{eval:imprecise}
\end{figure}

Class \texttt{ReferralReceiver} is a \texttt{BroadcastReceiver}, a
basic component in Android. In method 
\texttt{sendOtherBroadcastReceiver()}, an intent with action 
\texttt{com.android.vending.\linebreak INSTALL\_REFERRER} is created (line 3) and used to find all the \texttt{BroadcastReceivers} that can 
handle this intent (line 4). In lines 5 -- 9, appropriate
ones are created reflectively (lines 7 -- 8), 
on which \texttt{onReceive()} is called (line 9).

\ripple handles the incomplete information caused by \texttt{Context}, 
which contains some global information about an app. 
As \texttt{ctx = null}, \texttt{info.activityInfo.name = null}.
\ripple first applies
\rulename{C3-ForName} to \texttt{Class.forName(info.\linebreak 
activityInfo.name)} in line 7 and then \rulename{C12-New} to
\texttt{clz.\linebreak newInstance()} in line 8 to infer the reflectively
created objects from the cast \texttt{BroadcastReceiver}. As 
\texttt{BroadcastReceiver} is often extended, \texttt{bReceiver} is 
made to point to 17 different types of objects with 15 being false positives.

If \ripple works with an advanced intent analysis, these 15 false 
positives may be all avoided. By assuming optimistically that 
\texttt{ctx} points to the \texttt{Context} of the current app, 
and modeling the call \texttt{queryBroadcastReceivers()} to 
return the information about all the components that can handle
the intent \texttt{criterion},
\texttt{infoList} will just contain the information pertaining to the
three BroadcastReceivers,
\texttt{ReferralReceiver},
\texttt{GrowMobileInstallReceiver} and \texttt{Tracker}. By also
filtering out
\texttt{ReferralReceiver} path-sensitively in line 6, \ripple
will now be able to make \texttt{bReceiver} to point to only two
objects precisely, one object of type
\texttt{GrowMobileInstallReceiver} and one object of type
\texttt{Tracker}, in line 8.
A similar pattern also appears in the app \textit{Cooking Fever}.

\subsection{RQ3: Scalability}

Table~\ref{eval:rq2} compares the analysis times of \ripple,
\strinf, and \spark (\soot's pointer analysis without 
reflection analysis). For each app, the final harness that is
iteratively constructed by \FlowDroid (to
discover callbacks) is used as its \texttt{main} entry.
For all the 17 apps except two
(\texttt{canteenhd} and \texttt{angrybirds}), \ripple finishes 
in under 10 secs. For all the 17 apps combined, 
\spark, \strinf and \ripple spend a total of 82.5, 86.0 and 93.4 seconds, 
respectively. 

\subsection{RQ4: Security Analysis}

\Cref{eval:rq2} also compares \ripple, \strinf and \spark in terms
of their effectiveness for enabling \FlowDroid to find sensitive
data leaks in Android apps. For each analysis, \FlowDroid calls it
iteratively to build a harness for an app
(by modeling more and more callbacks discovered) until a fixed-point
is reached. \FlowDroid will then
perform a flow- and context-sensitive taint analysis on the 
inter-procedural CFG, which is constructed based on the call
graph (CG) that is computed for the app with respect to the final harness 
obtained. Thus, \FlowDroid can be precise, in practice.

For each app, as shown in \Cref{eval:rq2}, the number of data leaks,
together with sensitive source and sink calls, found by \FlowDroid
under \spark, \strinf and \ripple are compared. As discussed
in Section~\ref{sec:RQ1},
\ripple finds more true reflective targets than \strinf.
For each app, \ripple's CG is always a super-graph of \strinf's CG, 
which is always a super-graph of \spark's CG. In particular,
\ripple's CG is larger than \strinf's CG in
13 out of the 17 apps evaluated. As a result,
\FlowDroid detects a total of \num{1418}, \num{1486} and 
\num{1796} leaks under
\spark, \strinf and \ripple, respectively.

Let us examine \ripple and \strinf in more detail. For 
10 out of the 17 apps evaluated,
\FlowDroid reports the same number of leaks for each app under both
analyses. However, for the remaining seven apps,
\FlowDroid reports 310 more leaks under \ripple than \strinf
(highlighted by the
numbers in bold in the last column of \Cref{eval:rq2}). \ripple's
ability in finding more true reflective targets 
in these apps
than \strinf, as shown in \Cref{eval:rq1}, has certainly paid off.

Let us revisit two examples discussed in Section~\ref{sec:iie}
to understand reflection-induced privacy violations. Let us 
consider the code taken
from \textit{Angry Birds} in \Cref{motivating:intent}. 
Due to an undermined intent, \texttt{cName = null}. \ripple
infers a total of five reflectively created objects for 
\texttt{Class.newInstance()} in line 6 based on its post-dominant cast
\texttt{MMBaseActivity}. As discussed in Section~\ref{sec:ex},
all these five objects are true targets 
configured to provide different forms of advertisement, enabling
\FlowDroid to detect 49 leaks that are missed by \strinf, on the
methods called directly or indirectly on these objects. For this
entire app, \FlowDroid finds 54 more leaks under \ripple than \strinf.
A similar code pattern also appears in \textit{Dumb Ways to Die}
where \FlowDroid finds 77 more leaks under \ripple than \strinf.

Let us now consider the code snippet taken from 
\textit{Twist} in \Cref{motivating:library}. 
\ripple infers that
\texttt{i()},
\texttt{d()},
\texttt{w()},
\texttt{e()},
\texttt{v()} and
\texttt{wtf()} in class \texttt{Log}, where the last two are
false positives, are the potential targets invoked at
\texttt{logMtd.invoke(null, tag, msg)} in line 6.
Some sensitive data may be accidentally passed to \texttt{msg} and 
get written to log files, resulting in potential security vulnerabilities. 
Due to the six target methods discovered, \FlowDroid finds a total of
12 data leaks (= 2 sources $\times$ 6 sinks) from two sensitive sources, of 
which 4 from
\texttt{v()} and
\texttt{wtf()} are false positives.

\section{Related Work}
\label{sec:rel}

We review only the most relevant work on reflection analysis for
Android apps and Java programs.

\vspace*{-2ex}
\paragraph*{Android Apps} 

Ernst et al.~\cite{checker} presented \checker,
a data-flow analysis for Android apps with reflective
calls handled later \cite{sparta}.
As for the reflection resolution approach
used, \checker performs regular string inference
as \droidRA for constant class and method names but
requires user annotations to handle non-constant
class and method names. In contrast, \ripple aims to 
automatically infer reflection targets at reflective
calls according to the type information available.

Li et al.~\cite{droidra} introduced \droidRA, a string
inference analysis for resolving reflection in Android apps. 
In this work, the reflection resolution problem is 
reduced to one of solving a constant string propagation in
the program. In \droidRA, reflective calls can only be
resolved if their class and method names are constants
and ignored otherwise.

Rasthofer et al.~\cite{harvester} developed 
\textsc{Harvester},
an approach for automatically extracting runtime values 
from Android apps. \textsc{Harvester} takes an Android 
installation package and performs a backward slicing 
starting at a point of interest. Afterwards, a new, reduced
package is generated and executed on a stock Android 
emulator or real phone to log the values of interest at 
runtime, such as some class and method names that are 
dynamically loaded and invoked via reflection. 
\textsc{Harvester} is designed to fight against 
code obfuscation techniques in order to extract runtime 
values, such as class and method names,
that may be first encrypted and then used 
in reflection calls under some particular inputs. 
In contrast, \ripple attempts to infer reflective
targets used under these and other circumstances statically under all
possible inputs. It will be interesting to investigate
how to combine dynamic techniques such as \textsc{Harvester}
and static reflection analyses such as \ripple to make
a desired tradeoff among soundness, precision and scalability.

Zhauniarovich et al.~\cite{stadyna} introduced 
\textsc{Stadyna}, a system that interleaves static and 
dynamic analysis in order to reveal the program behaviors 
caused by dynamic code update techniques, such as 
dynamic class loading and reflection. \textsc{Stadyna}
requires a modified Android virtual machine to log 
the side-effects of program behaviors at runtime, 
including the targets accessed at reflective calls. This 
online analysis approach involves human efforts (e.g.,
in preparing for test inputs), but with no guarantee for
code coverage. In
contrast, \ripple is a static reflection analysis, 
enabling it to be integrated with
a range of static security analyses, such as \FlowDroid,
to achieve improved code coverage and soundness, as
demonstrated in this paper.

\vspace*{-2ex}
\paragraph*{Java Programs} 

There are several reflection analysis techniques
for Java 
programs~\cite{elf, solar, yannisreflection, livshits}. 
Earlier, Livshits et al.~\cite{livshits} suggested to
discover reflective targets by tracking the flow of
string constants representing class/method/field names
and infer reflective targets based on post-dominating
type casts for \texttt{Class.newInstance()} calls if
their class names are statically unknown strings.
Recently, Li. et al. introduced
\elf~\cite{elf} and \solar~\cite{solar} to apply
sophisticated type
inference to resolve reflective targets effectively. In
particular, 
\solar is able to accurately identify where reflection is resolved unsoundly or imprecisely. In addition, it
provides a mechanism to balance soundness,
precision and scalability, representing a state-of-the-art
solution for Java. 
A recent program slicing technique, called program 
tailoring~\cite{Li16}, can also be leveraged to resolve 
reflection calls precisely.
However, all the reflection analysis  techniques proposed for Java
cannot resolve a reflective call fully if the data-flows needed
(e.g., class or method names) at the call are null.

Reflection analysis usually works together with pointer analysis in order
to discover the targets at reflective calls. For 
pointer analysis techniques developed for Java programs, we refer to 
\cite{bddbddb,Lhotak06,Manu06,Ana02,Tian16,Yannis11,Shang12,Lu13,Shang12a,Nguyen05}.

\balance

\section{Conclusion}
\label{sec:conc}

Due to the ubiquity of mobile phones and the rapid development of other connected mobile devices (e.g., tablets and 
e-books), security vulnerabilities in Android apps, 
especially the presence of reflection, pose major
security threats. 
In this paper, we introduce a reflection analysis for
Android apps for discovering the behaviors of 
reflective calls, which can cause directly or indirectly 
security vulnerabilities such as privacy violations. 
We advance the state-of-the-art reflection analysis 
for Android apps, by 
(1) bringing forward the ubiquity of IIEs for static
analysis, (2) introducing \ripple, the first IIE-aware
reflection analysis, and
(3) demonstrating that \ripple can resolve reflection in 
real-world Android apps precisely and efficiently, and 
consequently,
improve the effectiveness of downstream Android security analyses.

In future work, we plan to combine \ripple with some
dynamic analysis and integrate \ripple with an advanced
pointer analysis to improve both soundness and precision.

\section*{Acknowledgement}

This research is supported by an Australian Research Council
grant, DP170103956.

\bibliographystyle{abbrv}
\bibliography{AndApp}

\end{document}